\documentclass[12pt,draftclsnofoot, onecolumn,a4paper]{IEEEtran}
\usepackage{graphicx,cite,epsfig,amssymb,amsmath}
\renewcommand{\QED}{\QEDopen}

\begin{document}
\markboth{IEEE Transaction on Automatic Control, Vol. XX,
No. Y, Month 2009} {}

\title{\mbox{}\vspace{1.00cm}\\
\textsc{Pareto Region Characterization for Rate Control in Multi-user systems and Nash Bargaining} \vspace{1.5cm}}

\author{\normalsize\vspace{1cm}
Zengmao Chen$^1$, Sergiy A. Vorobyov$^2$, Cheng-Xiang Wang$^1$, and John Thompson$^3$ \\
\vspace{1.8cm}
$^1$ Joint Research Institute for Signal and Image Processing\\
School of Engineering \& Physical Sciences\\
Heriot-Watt University\\
Edinburgh, EH14 4AS, UK.\\
Email:\{zc34, cheng-xiang.wang\}@hw.ac.uk\\
\vspace{0.3cm}
$^2$ Department of Electrical and Computer Engineering\\
University of Alberta\\
Edmonton, AB, T6G 2V4, Canada. \\
Email: vorobyov@ece.ualberta.ca\\
\vspace{0.3cm}
$^3$Joint Research Institute for Signal and Image Processing\\
Institute for Digital Communications\\
University of Edinburgh\\
Edinburgh, EH9 3JL,  UK.  \\
Email: john.thompson@ed.ac.uk
\thanks{This paper has been presented in part at IEEE ICC2009, Dresden, Germany, June 2009.}
}

\date{\today}
\renewcommand{\baselinestretch}{1.2}
\thispagestyle{empty}
\maketitle
\thispagestyle{empty}
\newpage
\setcounter{page}{1}

\begin{abstract}

In this paper, the problem of rate control in multi-user multiple-input multiple-output (MIMO) interference systems is formulated as a multicriteria optimization (MCO) problem. The Pareto rate region of the MCO problem is first characterized. It is shown that for the convexity of the Pareto rate region it is sufficient that the interference-plus-noise covariance matrices of multiple users with conflicting objectives approach identity matrix. The latter can be achieved by using either orthogonal signaling, tame-sharing, or interference cancellation strategies. It is argued that, in the case of high interference, the interference cancellation is preferable in order to increase the Pareto boundary and guarantee the convexity of the Pareto rate region. Secondly, the Nash bargaining (NB) is applied to transform the MCO problem into a single-objective problem. The characteristics of the NB over MIMO interference systems such as the uniqueness, the existence of the NB solution and the feasibility of the NB set are investigated. It is found that when the NB solution exists, the sufficient condition for the corresponding single-objective problem to have a unique solution is also that the interference-plus-noise covariance matrices of multiple users approach identity matrix. A simple multi-stage interference cancellation scheme, which leads to a larger convex Pareto rate region and, correspondingly, a unique NB solution with larger user rates compared to the orthogonal and time-sharing signaling schemes, is proposed. The convexity of the rate region, the effectiveness of the proposed interference cancellation technique, and the existence of the NB solution for MIMO interference systems are examined by means of numerical studies. The fairness of the NB solution is also demonstrated. Finally, the special cases of multi-input single-output (MISO) and single-input single-output (SISO) interference systems are also considered.

\end{abstract}

\begin{keywords}
\begin{center}
Multicriteria optimization, Pareto region, Nash bargaining, game theory, MIMO interference channel.
\end{center}
\end{keywords}

\newpage

\IEEEpeerreviewmaketitle
\vspace{-1cm}
\section{Introduction}
Multicriteria optimization (MCO), which is featured by the need to simultaneously optimize multiple noncommensurable or even incompatible objectives, has been widely used in automatic control, telecommunications, economics, and many other fields of engineering and science\cite{mco1}. An MCO problem usually admits infinite number of \textit{noninferior} \textit{solutions}, which form a \textit{Pareto} \textit{boundary} and encompass a \textit{Pareto} \textit{region} \cite{mco2}. A noninferior solution in the Pareto boundary is considered to be \textit{Pareto} \textit{optimal} in the sense that no other solution can improve the performance of some objectives without deteriorating other objective(s). The characteristics of the Pareto region are of particular interest in terms of shedding light on solving the corresponding MCO problem.

Two main processes are involved for solving the MCO problem, namely, computing and decision making \cite{mco3}. In the former process, a set of noninferior solutions for the MCO problem is obtained, while, in the latter one, a final solution among the noninferior solutions is determined according to certain preferences of the system under optimization. Depending on the principle according to how these two processes interact with each other in search for a compromise solution, the MCO problem can be solved using one of the following three approaches \cite{mco3}: \textit{priori}, \textit{posteriori} and \textit{progressive} articulation of preferences. As the name suggests, the decision maker expresses its preferences prior to the computing process in the first approach. These preferences are often given in the form of an utility function aggregating all objectives. As for the posteriori case, the preferences of the decision maker are not known before computing. Thus, all the noninferior solutions for the MCO problem should be found before decision making. Unlike the first two approaches, in the progressive articulation of preferences, computing and decision making are performed in an alternating manner. In each iteration, a population of noninferior solutions is maintained during computing; then, the decision maker chooses several suitable compromise solutions for the computing process in next iteration to produce a new generation of noninferior solutions. After rounds of iterations, the final solution is obtained. Evolutionary algorithms are usually employed in this approach~\cite{mco4}.

Among the three abovementioned approaches for solving the MCO problem, the priori articulation of preferences is most commonly used due to its simplicity and effectiveness. It can be further broken down into three broad classes of methods: \textit{scalarization\ of\ multiple\ objectives} - the MCO problem is transformed into a single-objective optimization problem by scalarizing the multiple objectives into a single-valued utility function \cite{mco3}; \textit{prioritization\ of\ multiple\ objectives} - a certain priority is assigned to each objective and the problem is optimized according to the priority order of its objectives \cite{mco5}; \textit{goal-value\ method} - a goal value (usually the minimum attained performance) is assigned to each objective before optimization to indicate its corresponding desired level of performance\cite{mco6}.

The rate control problem in multi-user interference systems can be formulated as an MCO problem. The convexity of the Pareto rate region of the MCO problem is a desirable feature for the interference system. Various approaches have been applied to convexify a rate region\footnote{When referring to rate region/boundary, we mean Pareto rate region/boundary hereafter, if there is no particular other clarification.} \cite{r4} \cite{r3}. These approaches can be broadly divided into two categories. The first one is based on the use of a convex hull (see, for example \cite{r4}), which corresponds to the so-called time-sharing strategies. However, the time-sharing rate boundary is achievable only in terms of average rate rather than instantaneous rate. The other approach is based on using orthogonal stategies (signaling schemes), e.g., time division multiplexing (TDM) and frequency division multiplexing (FDM) \cite{r3}. Orthogonal signaling schemes significantly simplify the MCO problem, but they usually lead to smaller achievable rate regions. The corresponding rate loss can be significant when the interference is low. Therefore, it is desirable to analyze the convexity of the true Pareto rate region only employing pure non-orthogonal strategies.

To solve the MCO problem in multi-user interference systems, several methods have been developed in the literature based on the priori articulation of preferences. The most representative one is sum-rate maximization, e.g., the sum-rate maximization for multi-user multiple-input multiple-output (MIMO) interference systems in \cite{r8}. The obvious drawback of this method is the lack of fairness, since the performance of users with bad channel conditions are always sacrificed. The fairness can be improved by weighted sum rate approaches like in \cite{wsum}, but it is not always easy to determine the weighted coefficients for all users. The preference could also be based on proportional fairness like in\cite{pfair}, where the fairness of a user is proportional to its channel conditions. Recently, game theory has been increasingly used in the control and optimization for communication systems like in \cite{smodular}. Specifically, the Nash bargaining (NB) from cooperative game theory has been widely used to solve the MCO problem for multi-user interference systems. NB is an effective scheme to balance the fairness of individual users and the system-level performance\cite{r11}. Representative examples in the literature include \cite{r3} and \cite{r5}, for single-input single-output (SISO), \cite{r4} and \cite{r6} for multiple-input single-output (MISO), and \cite{r7} for MIMO interference systems. In \cite{r7}, a practical suboptimal algorithm for finding the NB solution was designed by exploring the gradient projection method \cite{r8}. However, little research has been done to characterize the pure strategy based NB for MIMO interference systems.

In this paper, we apply MCO to rate control problem in multi-user interference systems. Specifically, we formulate the cooperative rate control of MIMO interference systems into an MCO problem. Firstly, the Pareto rate region of the MCO problem is characterized. It is proved that the interference-plus-noise covariance matrices approaching the identity matrix is a sufficient condition for the convexity of the rate region. Moreover, a significant implication is found that when interference is high, interference cancellation techniques are preferable in making the rate region convex. Secondly, the MCO problem for rate control in multi-user MIMO interference systems is converted to a single-objective Nash-product maximization problem using NB. The characteristics of the NB over MIMO interference systems such as the uniqueness of the NB solution, the feasibility of the NB set and the existence of the NB solution are studied. Furthermore, we propose a multi-stage interference cancellation scheme, which leads to a larger rate region and, correspondingly, an NB solution with larger rates compared to orthogonal and time-sharing signaling based approaches. More importantly, it also guarantees the convexity of the rate region when the interference is high, which consequently admits a unique NB solution.

The remainder of this paper is organized as follows. The MCO problem and Nash-product maximization problem for multi-user MIMO interference systems are formulated in Section~II. The main results including the characterization of both the Pareto rate region and the NB solution, and the elaboration for the proposed multi-user interference cancellation technique are given in Section III. The convexity of the rate region, effectiveness of the proposed interference cancellation technique, fairness, existence and uniqueness of the NB solution for MIMO interference systems are exemplified in Section IV via numerical studies. In Section V, some special cases are reviewed. Conclusions are drawn in Section VI.

\section{Problem Formulation}
\subsection{MCO in MIMO Interference Systems}

Consider an $M$-user MIMO interference system in which all users transmit on the same wireless channel simultaneously. The transmitter and receiver for each user are equipped with $N_{t}$ and $N_{r}$ antennas, respectively. The $N_{r}\times 1$ complex baseband signal vector  received by user $i\ (i=1,2,\ldots,M)$ can be written as \cite{r14}
\begin{equation} \label{eq1}
{\mathbf{y}}_i = \sqrt{\rho_i} {\mathbf{H}}_{ii} {\mathbf{x}}_i + \sum_{j = 1, j \neq i}^{M} \sqrt{\eta_{ij}} {\mathbf{H}}_{ij} {\mathbf{x}}_j + {\mathbf{n}}_i
\end{equation}
where $\rho_i$ is the normalized signal-to-noise ratio (SNR) for user $i$; $\eta_{ij}\ (i,j=1,\cdots,M,\ i\neq j)$ is the normalized interference-to-noise ratio (INR) from transmitter $j$ to receiver $i$; ${\mathbf{H}}_{ii}$ and  ${\mathbf{H}}_{ij}$ are the  $N_r \times N_t$  channel matrices from transmitter $i$ and transmitter $j$ to receiver $i$, respectively; ${\mathbf{x}}_i$ is the  $N_t \times 1$ transmitted signal vector for user $i$; and  ${\mathbf{n}}_i$ is the $N_r\times 1$ independent and identically distributed (i.i.d.) additive white Gaussian noise (AWGN) vector of user $i$ with zero mean and covariance matrix $E[{\mathbf{n}}_i {\mathbf{n}}_i^{H} ]=\mathbf{I}$. Here, $E[\cdot ]$ stands for the expectation operator,  $(\cdot )^H$ denotes the Hermitian transpose operation, and $\mathbf{I}$ is an $N_r\times N_r$ identity matrix.

We assume that:\hspace{-0.05cm} (i)\hspace{-0.05cm} each\hspace{-0.05cm} transmitter/receiver\hspace{-0.05cm} transmits/receives\hspace{-0.05cm} symbols\hspace{-0.05cm} independently;\hspace{-0.05cm} (ii)\hspace{-0.05cm} the\hspace{-0.05cm} co-channel interference from other users is unknown and treated as noise, i.e., no interference cancellation techniques are employed by receivers; (iii) the channel varies sufficiently slowly and is considered as time invariant during the period of each symbol transmission.

The mutual information for user $i$ can be expressed as \cite{r9}
\vspace{-0.1cm}
\begin{equation}\label{eq2}
I_i(\mathbf{Q}) = \log_2 \det \left({\mathbf{I}} + \rho_i {\mathbf{H}}_{ii} {\mathbf{Q}}_i {\mathbf{H}}_{ii}^H {\mathbf{R}}_{-i}^{-1} \right), \ \ i=1,\ldots,M
\end{equation}
where ${\mathbf{Q}}_i=E[{\mathbf{x}}_i{\mathbf{x}}_i^H]$ is the Hermitian positive semi-definite (PSD) transmit covariance matrix of the input signal vector for user $i$, i.e., $\mathbf{Q}_i\succeq 0$, and
\begin{equation}\label{eq3}
{\mathbf{R}}_{-i}={\mathbf{I}}+\sum_{j = 1, j\neq i}^{M} \eta_{ij} {\mathbf{H}}_{ij} {\mathbf{Q}}_{j} {\mathbf{H}}_{ij}^H,\ \ i=1,\ldots,M
\end{equation}
is the interference-plus-noise covariance matrix for user $i$. We define $\mathbf{Q}\triangleq ({\mathbf{Q}}_1,\ldots,{\mathbf{Q}}_L)$  as a set of transmit covariance matrices. Since the transmission of each user is power limited, the following trace constraint applies to ${\mathbf{Q}}_i$
\begin{equation}\label{eq4}
\mathrm{tr}({\mathbf{Q}}_i)\leq p_i,\ \ \ i=1,\ldots,M
\end{equation}
where $\mathrm{tr}( \cdot )$ denotes the trace operator. We also assume that each transmitter $i$ has the full knowledge of the channel, SNRs, INRs and transmit covariance matrices of all the other transmitters.

The rate control objective in the MIMO interference system is to maximize the rate\footnote{Hereafter, when referring to rate, we mean mutual information obtained using \eqref{eq2}.}of each user by optimizing their transmit covariance matrices ${\mathbf{Q}}_i\ (i=1,\ldots,M)$ under the trace constraints given by \eqref{eq4}. Therefore, rate control in the multi-user MIMO interference system can be formulated as the following MCO problem
\begin{align}\label{eq5}
\max_{\mathbf{Q}} I_i(\mathbf{Q}) \ \ \ &i=1,\ldots,M\nonumber\\
{\rm subject\ to}\ \ \ &\mathbf{Q}_i\succeq 0, \ \ i=1,\ldots,M\nonumber \\
&\mathrm{tr}(\mathbf{Q}_i)\leq p_i, \ \ i=1,\ldots,M.
\end{align}

\subsection{Scalarization of the MCO using NB}

According to game theory, a game consists of three elements: \textit{players}, \textit{strategy} and \textit{utility}~\cite{game1}. Players are rational parties involved in the game. Strategy stands for actions or behaviors taken by players. Utility is usually defined in the form of a certain performance metric for players. The MIMO interference system delineated above can be modeled as a MIMO interference game, whose players are the users in the MIMO system. The rate of each user represents the utility of the corresponding player. The transmit covariance matrix of each user forms the strategy space of each player.

A game can be classified as either competitive or cooperative according to the cooperation scheme among players. In a competitive game, as the name suggests, all the players compete with each other selfishly and rationally. Players neither communicate nor cooperate with each other during the game. A steady state in a competitive game for which each player can not improve its utility by unilaterally changing its own strategy is called the NE \cite{game1}.
For a MIMO interference game, the NE can be mathematically expressed as
\begin{equation}\label{eq6}
\forall i=1,\ldots,M,\ \mathbf{Q}_i\succeq 0,\ \mathrm{tr}({\mathbf{Q}}_i)\leq p_i : \ I_i({\mathbf{Q}}_i^*,{\mathbf{Q}}_{-i}^*) \geq I_i({\mathbf{Q}}_i,{\mathbf{Q}}_{-i}^*)
\end{equation}
where ${\mathbf{Q}}_i^*$ and ${\mathbf{Q}}_{-i}^*$ denote the transmit covariance matrices of the NE for user $i$ and for all the other users except $i$, respectively.
The transmit covariance matrix of each player leading to the NE can be found via iterative water filling (IWF) as \cite{r10}
\begin{equation}\label{eq7}
{\mathbf{Q}}_i^*={\mathbf{U}}_i (\mu_i \mathbf{I} - \mathbf{D}_i^{-1})^+ {\mathbf{U}}_i^H,\ \ \ i=1,\ldots,M
\end{equation}
where ${\mathbf{U}}_i {\mathbf{D}}_i {\mathbf{U}}_i^H=\rho_i{\mathbf{H}}_{ii}^H {\mathbf{R}}_{-i}^{-1} {\mathbf{H}}_{ii}$ is the eigendecomposition of $\rho_i{\mathbf{H}}_{ii}^H {\mathbf{R}}_{-i}^{-1} {\mathbf{H}}_{ii}$, ${\mathbf{U}}_i\ (i=1,\ldots,M)$ is the unitary matrix of eigenvectors, ${\mathbf{D}}_i\ (i=1,\ldots,M)$ is a diagonal matrix of eigenvalues, and $\mu_i$ denotes the power level given by IWF.

Generally, the NE is not optimal from the system point of view due to its competitive and selfish nature. Whereas, in a cooperative game, all the players negotiate with each other prior to the game, which usually results in utility improvement\cite{nash} \cite{game2}. There exist many cooperative game-theoretic approaches. In this paper, we restrict our attention to the NB as the simplest and also the most popular one \cite{r11} to scalarize the multiple objectives of the MCO problem \eqref{eq5}.

The NB is well defined in a convex rate region \cite{r11}. In the context of the MIMO interference game, for the case when the rate region is convex, the bargaining set, i.e., the set of available strategies for user $i$, can be expressed as
\begin{equation}\label{eq8}
S=\{ \mathbf{Q}_i | \mathbf{Q}_i\succeq 0,\ \mathrm{tr}(\mathbf{Q}_i) \leq p_i,\  {\rm and}\ I_i(\mathbf{Q}) > I_i^{N\hspace{-0.06cm}E} ,i = 1, \ldots, M \}
\end{equation}
where $I_i^{N\hspace{-0.06cm}E}$ is the utility of the NE for user $i$. It also has a \textit{disagreement\ point}, which is defined as the state that players resort to when the cooperation fails. Usually, the NE is taken as the disagreement point in NB. By applying the NB, the multiple objectives in the MCO problem \eqref{eq5} can be scalarized. Then, the MCO problem can be transformed into the following single-objective optimization problem
\begin{align}\label{eq9}
\max_{\mathbf{Q}} \prod _{i=1}^{M} (I_i(\mathbf{Q}&) - I_i^{N\hspace{-0.06cm} E})\nonumber\\
{\rm subject\ to}\ \ \ &\mathbf{Q}_i\succeq 0, \ \ i=1,\ldots,M\nonumber \\
&\mathrm{tr}(\mathbf{Q}_i)\leq p_i, \ \ i=1,\ldots,M\nonumber \\
&I_i(\mathbf{Q})>I_i^{N\hspace{-0.06cm} E}, \ \ i=1,\ldots,M \ .
\end{align}
It is worth noting that the NB corresponds to \textit{proportional\ fairness} and the Nash product $\prod _{i=1}^{M} (I_i(\mathbf{Q}) - I_i^{N\hspace{-0.06cm} E})$ is converted to the rate product of all users when $I_i^{N\hspace{-0.06cm} E}=0$ for all $i$. An intuitive explanation of the above optimization problem is that the NB introduces a cooperation scheme among all MIMO users and regulates their transmissions (transmit covariance matrices) under the power constraint. On one hand, it guarantees that the utility of each MIMO user is not less than the one given by the NE. On the other hand, it maximizes the Nash product of the whole MIMO system. Therefore, it provides a good trade off between the fairness requirements to individual users and the overall performance of the whole system.

In the sequel, we first characterize the Pareto rate region of the MIMO interference systems. Then, we study several characteristics of the NB in MIMO interference systems such as the uniqueness of the NB solution, the feasibility of the NB set and the existence of the NB solution. Finally, a simple multi-stage interference cancellation scheme is proposed to convexify the rate region.

\section{Main Results}

\subsection{Characterization of Pareto Rate Region}
The convexity of the rate region is a desirable feature for communication systems. Its immediate merit is that it usually yields larger rate region compared to a nonconvex one. More importantly, the convexity of the rate region is a necessary condition to ensure the convexity of the MCO problem. It simplifies the MCO problem and makes it solvable and practically realizable. There exist two broad approaches to convexify a rate region. The most common and direct approach is based on the use of a convex hull, which corresponds to the time-sharing signaling. In this case, all users agree to use a set of transmitting strategies for certain fraction of time and to use another set of strategies during the rest of the time (see, for example \cite{r4}). However, time-sharing signaling leads to a convex rate region in terms of the average rate (over time). Thus, the time-sharing rate boundary may not be achievable in terms of instantaneous rate~\cite{r4}.

Another approach is based on the use of orthogonal signaling among cooperative users. The examples of orthogonal signaling are TDM\footnote{TDM distinguishes itself from the time-sharing signaling/convex hull in the sense that for the 2-user case the TDM rate boundary is a line connecting two single-user points, while time-sharing signaling corresponds to the outmost lines between points in the Pareto rate boundary as shown in Fig.~2.} and FDM. In this case, users agree to split the degrees of freedom (time or bandwidth) into several orthogonal parts and each user uses one part only. The orthogonal signaling significantly simplifies the MCO problem, but the associated rate loss may be significant, especially when interference is low. Instead of using either time-sharing or orthogonal signaling, we investigate the convexity of the true Pareto rate region only using pure strategies. The true Pareto rate region is both achievable in terms of instantaneous rate and usually larger than the orthogonal signaling-based counterpart. In the context of the MIMO interference systems, we derive the following sufficient condition ensuring the convexity of its rate region.

$\textbf{Proposition\ \hspace{-0.1cm}1}:$ {\it If the
interference-plus-noise covariance matrices $\mathbf{R}_{-i} \to
\mathbf{I}\ \ (i=1,\cdots,M)$, the Pareto rate region of the MIMO
interference system is convex. }

\begin{proof}
It is straightforward to prove that the rate region is convex when the utility functions of all users $I_i(\mathbf{Q})\ (i=1,\cdots,M)$ in \eqref{eq2} are concave functions of $\mathbf{Q}$. Let us prove the concavity of the function $I_i(\mathbf{Q})$, $i=1,\ldots,M$ under the condition that interference-plus-noise covariance matrices approach $\mathbf{I}$. Note that a function $g(x)$ is concave if and only if (i) $f(t)=g(tx_1+(1-t)x_2),\ 0\leq t \leq 1$ is a concave function of $t$ for any feasible $x_1$ and $x_2$, which is equivalent to $f''(t)=d^2f(t)/dt^2\leq 0$; and (ii) the domain of $g(x)$ is convex \cite{r12}.

We consider the following convex combination of two different sets of transmit covariance matrices: $(\mathbf{X}_1,\ldots,\mathbf{X}_L)$ and $(\mathbf{Z}_1,\ldots,\mathbf{Z}_L)$, that is
\begin{eqnarray} \label{eq10}
\mathbf{Q}(t)
&=&(1-t)(\mathbf{X}_1,\ldots,\mathbf{X}_M) + t(\mathbf{Z}_1,\ldots,\mathbf{Z}_M)\nonumber \\
&=&(\mathbf{X}_1,\ldots,\mathbf{X}_M) + t(\mathbf{Z}_1 -\mathbf{X}_1,\ldots,\mathbf{Z}_M - \mathbf{X}_M)\nonumber \\
&=&(\mathbf{X}_1,\ldots,\mathbf{X}_M) + t(\mathbf{Y}_1,\ldots,\mathbf{Y}_M)
\end{eqnarray}
where $0\leq t \leq 1$ and $\mathbf{Y}_i=\mathbf{Z}_i -\mathbf{X}_i\ (i=1,\ldots,M)$. Let us expand the utility function \eqref{eq2} as
\begin{eqnarray}\label{eq11}
f_i(t) \!\!&=&\!\!I_i(\mathbf{Q}(t))\nonumber \\
\!\!&=&\!\! \log_2  \det \left({\mathbf{I}} +
\rho_i {\mathbf{H}}_{ii} {\mathbf{Q}}_i
{\mathbf{H}}_{ii}^H {\mathbf{R}}_{-i}^{-1}
\right) \nonumber \\
\!\!&=&\!\! \frac{1}{\ln 2} \ln \frac{\det (\mathbf{R}_{-i}+\rho_i
\mathbf{H}_{ii} \mathbf{Q}_{i} \mathbf{H}_{ii}^H)}{\det
(\mathbf{R}_{-i})},\ \ i=1, \ldots,M.
\end{eqnarray}
Applying the well known property of matrix differential calculus \cite{r13}, i.e.,
\begin{equation}\label{eq12}
\frac{d}{dx} \ln \det \left(\mathbf{A}(x)\right) = \mathrm{tr} \left(\mathbf{A}(x)^{-1}\frac{d\mathbf{A}(x)}{dx}\right),
\end{equation}
we can obtain the first derivative of $f_i(t)$ as
\begin{eqnarray}\label{eq13}
f_i^{'}( t) \!\!\!&=&\!\!\! \frac{1}{\ln 2} \left[ \frac{d}{dx}
\ln \det (\mathbf{R}_{-i}+\rho_i \mathbf{H}_{ii} \mathbf{Q}_{i}
\mathbf{H}_{ii}^H) - \frac{d}{dx} \ln \det
(\mathbf{R}_{-i}) \right] \nonumber \\
\!\!\!&=&\!\!\! \frac{1}{\ln 2} \left[\mathrm{tr} \! \left( \left(
{\mathbf{R}_{-i}} \!+\! \rho_i \mathbf{H}_{ii} \mathbf{Q}_{i}
\mathbf{H}_{ii}^H \right)^{-1} \! \left(
\frac{d\mathbf{R}_{-i}}{dt} \!+\! \rho_i \mathbf{H}_{ii}
\mathbf{Y}_{i}  \mathbf{H}_{ii}^H \right) \! \right) \!-\! \mathrm{tr}
\!\left( \mathbf{R}_{-i}^{-1} \frac{d\mathbf{R}_{-i}}{dt} \right)
\!\right] \!\!.
\end{eqnarray}
In \eqref{eq12}, $\mathbf{A}(x)$ is a matrix function of scalar parameter $x$. Using \eqref{eq13} and applying two other properties of matrix differential calculus \cite{r13}, i.e.,
\begin{equation}\label{eq14}
\frac{d}{dx} \mathrm{tr}\left(\mathbf{A}(x)\right) = \mathrm{tr} \left(\frac{d\mathbf{A}(x)}{dx}\right)
\end{equation}
\begin{equation}\label{eq15}
\frac{d}{dx} \mathbf{A}(x)^{-1}=-\mathbf{A}(x)^{-1}\frac{d\mathbf{A}(x)}{dx}\mathbf{A}(x)^{-1},
\end{equation}
the second derivative of $f_i(t)$ can be expressed as
\begin{eqnarray}\label{eq16}
f_i^{''}(t) \!\!&=&\!\! \frac{1}{\ln 2} \left[\mathrm{tr} \left( - \left(
\mathbf{R}_{-i} + \mathbf{M}_i \right)^{-1}  \left(
\frac{d\mathbf{R}_{-i}}{dt} + \mathbf{N}_i \right) \left(
\mathbf{R}_{-i} + \mathbf{M}_i \right)^{-1} \left(
\frac{d\mathbf{R}_{-i}}{dt} + \mathbf{N}_i
\right) \right) \right. \nonumber \\
\!\!& &\!\! \qquad + \left. \mathrm{tr} \left( \mathbf{R}_{-i}^{-1}
\frac{d\mathbf{R}_{-i}}{dt} \mathbf{R}_{-i}^{-1}
\frac{d\mathbf{R}_{-i}}{dt} \right)\right]
\end{eqnarray}
where $\mathbf{M}_i=\rho_i \mathbf{H}_{ii} \mathbf{Q}_i \mathbf{H}_{ii}^H$ and $\mathbf{N}_i=\rho_i \mathbf{H}_{ii} \mathbf{Y}_i \mathbf{H}_{ii}^H$.

Let $\mathbf{A}_i=\left( {\mathbf{R}_{-i}} +  \mathbf{M}_{i} \right)^{-1}$ and $\mathbf{B}_i=d\mathbf{R}_{-i}/dt + \mathbf{N}_{i}$. Since $\mathbf{A}_i\succeq 0$, there exists a matrix $\mathbf{C}_i$ such that $\mathbf{A}_i=\mathbf{C}_i\mathbf{C}_i^H$ . Thus, the first trace in the right hand side of \eqref{eq16} can be written as
\begin{eqnarray}\label{eq17}
\mathrm{tr}\left( - \mathbf{A}_i \mathbf{B}_i \mathbf{A}_i \mathbf{B}_i
\right) \!\!&=&\!\!-\mathrm{tr}\left( \mathbf{C}_i \mathbf{C}_i^H
\mathbf{B}_i \mathbf{C}_i \mathbf{C}_i^H  \mathbf{B}_i \right)\nonumber \\
\!\!&=&\!\!-\mathrm{tr}\left( \mathbf{C}_i^H \mathbf{B}_i \mathbf{C}_i
\mathbf{C}_i^H  \mathbf{B}_i \mathbf{C}_i \right) \nonumber \\
\!\!&=&\!\!-\mathrm{tr}\left( \left( \mathbf{C}_i^H \mathbf{B}_i
\mathbf{C}_i \right) \left( \mathbf{C}_i^H  \mathbf{B}_i
\mathbf{C}_i \right)^H \right) \leq 0
\end{eqnarray}
The last equality in \eqref{eq17} is obtained using the fact that $\mathbf{B}_i$ is Hermitian. The inequality in \eqref{eq17} holds due to the fact that $\left( \mathbf{C}_i^H \mathbf{B}_i \mathbf{C}_i \right) \left( \mathbf{C}_i^H  \mathbf{B}_i \mathbf{C}_i \right)^H\succeq 0$.

When $\mathbf{R}_{-i} \to \mathbf{I}$, in the right hand side of \eqref{eq16},  $d\mathbf{R}_{-i}/dt=\sum_{j\neq i} \eta_{ij} {\mathbf{H}}_{ij} {\mathbf{Y}}_{ij} {\mathbf{H}}_{ij}^H$ approaches $\mathbf{0}$. Then, in the first trace of \eqref{eq16}, $( \mathbf{R}_{-i} \hspace{-0.08cm}+ \hspace{-0.08cm}\mathbf{M}_i \hspace{-0.05cm})^{-1}$ and $(d\mathbf{R}_{-i}/dt + \mathbf{N}_i)$ are dominated by $(\mathbf{I}+\mathbf{M}_i)^{-1}$ and $\mathbf{N}_i$, respectively. Note also that the second trace can be ignored as compared to the first one. Therefore, when $\mathbf{R}_{-i} \to \mathbf{I}$, $f_i^{''}(t)\leq 0$.

The domain of the utility function $I_i(\mathbf{Q})$ for the MIMO interference system is $\{ \mathbf{Q}_i | \mathbf{Q}_i\succeq 0,\ \mathrm{tr}(\mathbf{Q}_i) - p_i  \leq 0,\ i=1,\ldots,M \}$, which is clearly convex. Therefore, the utility function $I_i(\mathbf{Q})$ is concave and consequently the rate region is convex when $\mathbf{R}_{-i} \to \mathbf{I}$.
\end{proof}

The condition that $\mathbf{R}_{-i} \to \mathbf{I}$ means that $\sum_{j\neq i} \eta_{ij} {\mathbf{H}}_{ij} {\mathbf{Y}}_{ij} {\mathbf{H}}_{ij}^H  \to \mathbf{0}$. This corresponds to the condition that INRs $\eta_{ij}\ (i,j=1,\cdots,M,\ i\neq j)$ in \eqref{eq1} are sufficiently small. Hence, the immediate implication of Proposition 1 is that the rate region is convex when the interference is low.

In summary, we know that the orthogonal signaling is a sufficient condition for convexity of the rate region\cite{r3}. At the same time, Proposition~1 suggests that strict signal orthogonality is not a necessary condition for convexity of the rate region. Another sufficient condition which guarantees the convexity of the rate region is $\mathbf{R}_{-i} \to \mathbf{I}\ \ (i=1,\cdots,M)$. It generalizes the requirements of the time-sharing and orthogonal signaling, when constructing a convex rate region. More importantly, the following remark of a practical significance can be made.

$\textbf{Remark\ 1}$: {\it When the interference is high,
interference cancellation techniques outperform orthogonal
signaling techniques for convexifying a rate region in the sense that they lead to a convex rate region of a larger~size. }

Orthogonal signaling is a simple and widely-used method to produce a convex rate region at the cost of rate loss.  When the interference is high, applying interference cancellation techniques to a MIMO interference system eventually transforms a high-interference system into a low-interference system, which consequently leads to a convex rate region according to the Proposition~1 and results in a larger rate region compared to the true Pareto rate region. Therefore, interference cancellation techniques outperform orthogonal signaling techniques in terms of the size of the resultant rate region.

Various interference cancellation techniques \cite{cr_ic} can be applicable to interference systems. The widely used multi-user detection technique\cite{pic}, including sucessive interference cancellation (SIC) and parallel interference cancellation (PIC), is one suitable candidate for interference systems. Specifically, for MIMO interference systems, beamforming is another effective interference rejection technique. Transmit and receive beamforming can be performed in MIMO transmitters and receivers to mitigate interference, respectively. However, providing the interference cancellation capabilities for the interference system requires increase of both computational complexity and communication overhead among users. Take SIC and PIC for instance, these two approaches need to decode and reconstruct the interfering signals before substracting them from the received signal. The decoding and reconstruction processes definitely complicate the intereference system. Besides, channel information for the interfering channels is also required by the decoding and reconstruction. Moreover, additional communication overhead among receivers is necessary as well when receiver cooperation is employed to further enhance the interference cancellation. The additional complexity and communication overhead are paid off by the increase of rate region.

\subsection{Characterization of NB Solution}

\textit{1)\ Uniqueness\ of\ NB\ Solution:}

It is straightforwad to see that the optimization problem given by
\eqref{eq9} is identical to
\[ \hspace{-1.7cm}\max_{\mathbf{Q}}\
\ln\left(\ln^L 2 \prod _{i=1}^{M} \left( I_i(\mathbf{Q}) - I_i^{N
\! E}\right)\right)
\]
\vspace{-1.3cm}

\begin{align}\label{eq18}
{\rm subject\ to}\ \ \ &\mathbf{Q}_i\succeq 0, \ \ i=1,\ldots,M\nonumber \\
&\mathrm{tr}(\mathbf{Q}_i) - p_i\leq 0, \ \ i=1,\ldots,M\nonumber \\
&I_i^{N\hspace{-0.06cm} E} -I_i(\mathbf{Q})< 0, \ \ i=1,\ldots,M \ .
\end{align}
Note that \eqref{eq18} is a convex optimization problem if and only if its objective function is concave and its constraint set is convex \cite{r12}.


$\textbf{Proposition\ 2}:$ {\it When the interference-plus-noise
covariance matrices $\mathbf{R}_{-i} \to \mathbf{I}$ $(i=1,\cdots,M)$ , the optimization problem \eqref{eq18} is a convex
problem. }

\begin{proof}
First, by using the same methodology as the one used to prove
Proposition~1, let us show that the objective function in
\eqref{eq18} is concave under the condition that the
interference-plus-noise covariance matrices $\mathbf{R}_{-i}$
approach $\mathbf{I}$, $\forall i$. Considering the convex
combination in \eqref{eq10}, the objective function of
\eqref{eq18} can be expanded using \eqref{eq2} as
{ \[ \hspace{-2.75cm}f(t)=\ln \left(\ln^L 2 \prod_{i=1}^{M}
(I_i(\mathbf{Q}(t)) - I_i^{N\hspace{-0.06cm} E}) \right)
\]}
\vspace{-0.5cm}
\begin{eqnarray}\label{eq19}
\hspace{0.77cm}\ \ \ \ =\sum_{i=1}^L \ln\left(\ln \frac{\det
(\mathbf{R}_{-i}+\rho_i \mathbf{H}_{ii} \mathbf{Q}_{i}
\mathbf{H}_{ii}^H)}{\det (\mathbf{R}_{-i})} - \ln 2 \!
I_i^{N\hspace{-0.06cm} E}\right).
\end{eqnarray}
Let us define
\vspace{-0.25cm}
\begin{equation}\label{eq20}
T_i=\ln \frac{\det (\mathbf{R}_{-i}+\rho_i \mathbf{H}_{ii}
\mathbf{Q}_{i} \mathbf{H}_{ii}^H)}{\det (\mathbf{R}_{-i})} - \ln 2
I_i^{N\hspace{-0.06cm} E}.
\end{equation}
Using \eqref{eq12}, \eqref{eq14} and \eqref{eq15}, we obtain the second derivative of $f(t)$ as
\vspace{-0.15cm}
\begin{equation}\label{eq21}
f^{''}(t)=\sum_{i=1}^M \alpha + \beta + \gamma
\end{equation}
\vspace{-0.3cm}
where
\begin{eqnarray}
\alpha \!\!&=&\!\! \frac{\mathrm{tr}\left( -  \left( \mathbf{R}_{-i} +
\mathbf{M}_i \right)^{-1}  \left( \frac{d\mathbf{R}_{-i}}{dt}
+ \mathbf{N}_i \right)  \left( \mathbf{R}_{-i} + \mathbf{M}_i
\right)^{-1}  \left( \frac{d\mathbf{R}_{-i}}{dt} + \mathbf{N}_i
\right)  \right)}{T_i} \label{eq22} \\
\beta \!\!&=&\!\! \frac{\mathrm{tr}\left( \mathbf{R}_{-i}^{-1}
\frac{d\mathbf{R}_{-i}}{dt} \mathbf{R}_{-i}^{-1}
\frac{d\mathbf{R}_{-i}}{dt} \right)}{T_i} \label{eq23}\\
\gamma \!\!&=&\!\! - \frac{\left[ \mathrm{tr} \left( \left(
{\mathbf{R}_{-i}} + \mathbf{M}_{i} \right)^{-1} \left(
\frac{d\mathbf{R}_{-i}}{dt} + \mathbf{N}_{i} \right) \right)
\hspace{-0.08cm} -\mathrm{tr} \hspace{-0.08cm} \left( \mathbf{R}_{-i}^{-1}
\frac{d\mathbf{R}_{-i}}{dt} \right) \right]^2}{T_i^2}.\label{eq24}
\end{eqnarray}
Similar to \eqref{eq17}, the numerator of $\alpha$ in \eqref{eq22} is not positive. If the bargaining set $S$  is not empty, then $T_i >0$ . Thus, $\alpha$ is not positive. Similarily, $\beta$ in \eqref{eq23} can be ignored as compared to $\alpha$ when the interference-plus-noise covariance matrices approach $\mathbf{I}$. One can also see that $\gamma$ in \eqref{eq24} is not positive. Therefore, when the interference-plus-noise covariance matrices $\mathbf{R}_{-i}$ approach $\mathbf{I}$, $f^{''}(t)\leq 0$.

The domain of the objective function in \eqref{eq18} is $\{ \mathbf{Q}_i |
\mathbf{Q}_i\succeq 0,\ \mathrm{tr}(\mathbf{Q}_i) - p_i  \leq 0,\
i=1,\ldots,M \}$, which is obviously convex. Therefore, the
objective function in \eqref{eq18} is concave.

As a next step, let us prove the convexity of the constraint set.
The constraint set of \eqref{eq18} is identical to the bargain set \eqref{eq8}.
Specifically, it can be rewritten as
\begin{eqnarray}\label{eq25}
S \!\!\!&=&\!\!\! \{ \mathbf{Q}_i \, | \, \mathbf{Q}_i\succeq 0,
\, \mathrm{tr}(\mathbf{Q}_i) \!-\! p_i \!\leq\! 0,\, i \!=\! 1,\ldots,M \}
\cap \{ \mathbf{Q}_i \, | \, -\! I_i(\mathbf{Q}) \!+\! I_i^{N \!
E} \!\leq\! 0, \, i \!=\! 1,\ldots,M \}
\nonumber \\
&=&S_1 \cap S_2.
\end{eqnarray}
It is easy to establish the convexity of the subset $S_1$ in \eqref{eq25}, but the convexity of the subset $S_2$ is not obvious. Let us define $h(t)=-I_i(\mathbf{Q}(t))+I_i^{N\hspace{-0.06cm} E}$. Adopting the same methodology as the proof for the concavity of the utility function $I_i(\mathbf{Q})$, we have
\begin{eqnarray}\label{eq26}
h^{''} (t) \!\!&=&\!\! \mathrm{tr}\left( \left( \mathbf{R}_{-i} +
\mathbf{M}_i \right)^{-1} \left( \frac{d\mathbf{R}_{-i}}{dt} +
\mathbf{N}_i \right)  \left( \mathbf{R}_{-i} + \mathbf{M}_i
\right)^{-1} \left( \frac{d\mathbf{R}_{-i}}{dt} + \mathbf{N}_i
\right) \right. \nonumber \\
\!\!& &\!\! \qquad \left. - \mathbf{R}_{-i}^{-1}
\frac{d\mathbf{R}_{-i}}{dt} \mathbf{R}_{-i}^{-1}
\frac{d\mathbf{R}_{-i}}{dt} \right).
\end{eqnarray}
A similar result can be obtained that when the interference-plus-noise covariance matrices $\mathbf{R}_{-i}$ approach $\mathbf{I}$, the second term inside the trace operator in \eqref{eq26} can be ignored when compared to the first one. Thus, similar to \eqref{eq17}, if the interference-plus-noise covariance matrices $\mathbf{R}_{-i}$ approach $\mathbf{I}$, \eqref{eq26} can be rewritten as
\begin{eqnarray}\label{eq27}
h^{''}(t)&\approx& \mathrm{tr}\left( \left( \mathbf{R}_{-i} + \mathbf{M}_i \right)^{-1}  \left( \frac{d\mathbf{R}_{-i}}{dt} + \mathbf{N}_i \right)  \left( \mathbf{R}_{-i} + \mathbf{M}_i \right)^{-1}  \left(\frac{d\mathbf{R}_{-i}}{dt} + \mathbf{N}_i \right) \right) \nonumber \\
&=&\mathrm{tr}\left( \left( \mathbf{C}_i^H \mathbf{B}_i \mathbf{C}_i \right) \left( \mathbf{C}_i^H  \mathbf{B}_i \mathbf{C}_i \right)^H \right) \geq 0
\end{eqnarray}
which implies that when the interference-plus-noise covariance matrices $\mathbf{R}_{-i}$ approach $\mathbf{I}$, $h(t)$ is convex \cite{r12}, i.e., the subset $S_2$ in \eqref{eq25} is convex. Consequently, the constraint set $S$ in \eqref{eq25} is convex as well.

As we can see, the condition that the interference-plus-noise covariance matrices $\mathbf{R}_{-i}$ approach $\mathbf{I}$ is sufficient for both the concavity of the objective function in \eqref{eq18} and the convexity of its constraint set. Therefore, the Proposition 2 is proved.
\end{proof}

From Proposition 2, the following proposition can be obtained.

$\textbf{Proposition\ 3}$: {\it When the NB solution exists, the
condition that the interference-plus-noise covariance matrices
$\mathbf{R}_{-i}$ approach $\mathbf{I}$, is the sufficient
condition for the uniqueness of the NB solution of MIMO
interference systems. }

\begin{proof}
The interference-plus-noise  covariance matrices $\mathbf{R}_{-i}$
approaching $\mathbf{I}$ can ensure the convexity of the rate
region (see Proposition~1) on which the NB is defined.
Interestingly, it also guarantees the uniqueness of the NE in MIMO
interference systems (see Proposition~1 in \cite{part1}), i.e., it
guarantees the IWF convergence. Moreover, Proposition~2 states
that if the interference-plus-noise covariance matrices
$\mathbf{R}_{-i}$ approach $\mathbf{I}$, then \eqref{eq18} is a convex
optimization problem, i.e., there exists at most one solution
maximizing the objective function of \eqref{eq18} \cite{r12}. Therefore,
the requirement of the rate region convexity is the convexity of
the optimization problem \eqref{eq18}. Moreover, the existence of the
solution for problem \eqref{eq18} is the sufficient condition for the
uniqueness of the NB solution. Applying Propositions~1~and~2 in
this paper and Proposition~1 in \cite{part1}, we complete the
proof.
\end{proof}

For proportional fairness, where the Nash product corresponds
to the rate product, it is easy to infer from \eqref{eq18} that the
interference-plus-noise covariance matrices $\mathbf{R}_{-i}$
approaching $\mathbf{I}$ is also the sufficient condition for the
concavity of $\ln(\ln^L 2 \prod _{i=1}^{M} I_i(\mathbf{Q}))$,
which is equivalent to the concavity of the rate product logarithm
for MIMO interference systems.


\textit{2)\ Existence of NB Solution and Feasibility of NB Set:}

As stated in Section II, in MIMO interference systems, the NE is a
steady and optimal state for selfish and competitive users
competing with each other. However, the NE is not Pareto
efficient, i.e., it is usually below the rate boundary. If all the
users simultaneously deviate from the NE and agree to adopt
different strategies (i.e., transmit covariance matrices), this
may result in simultaneous rate improvement for all users. In this
case, the NB solution will exist.

The existence of the NB solution highly depends on the underlying MIMO interference channel, including its channel matrices, SNRs and INRs. In general, it is difficult to find a closed-form expression or quantitive description specifying when the NB solution exists. In the next section, we approach this problem by numerical studies and examine the impact of SNR and INR on the existence of the NB solution.

Another important characteristic of the NB is its feasible bargaining set. Some general characteristics of the feasible utility set have been studied under an axiomatic framework in \cite{structure}. In what follows, we will analyze the structure of the transmit covariance matrices which form the feasible bargaining set of the NB in MIMO interference systems without employing time-sharing or orthogonal signaling. The transmit covariance matrix $\mathbf{Q}_i\ (i=1,\cdots,M)$ of each user is Hermitian PSD. Thus, it can be decomposed as
\begin{equation}\label{eq28}
\mathbf{Q}_i=\mathbf{V}_i \mathbf{\Lambda}_i \mathbf{V}_i^H,\ \ \ i=1,\cdots,M
\end{equation}
where $\mathbf{V}_i\ (i=1,\cdots,M)$ are unitary matrices of eigenvectors of $\mathbf{Q}_i\ (i=1,\cdots,M)$ and $\mathbf{\Lambda}_i\ (i=1,\cdots,M)$ are the diagonal matrices with the corresponding eigenvalues. The transmit covariance matrices in the feasible set~\eqref{eq8} can be considered as evolving from transmit covariance matrices of the NE~\eqref{eq7}. The evolution can be performed in three different ways: (i)  changing the diagonal matrix $(\mu_i \mathbf{I} - \mathbf{D}_i^{-1})^+$ of \eqref{eq7}; (ii)    modifying the unitary matrix ${\mathbf{U}}_i$ of \eqref{eq7}; (iii) changing both the diagonal and unitary matrices of \eqref{eq7}. The feasible set $S$ in \eqref{eq25} can be comprised of transmit covariance matrices of the following possible types:\renewcommand{\labelenumi}{\Roman{enumi}.}
\begin{itemize}
\item\textit{Type I:} The transmit covariance matrices $\mathbf{Q}_i$ have the same unitary matrices $\mathbf{V}_i$, but different diagonal matrices $\mathbf{\Lambda}_i$ from $\mathbf{Q}_i^*$.\hspace{-0.05cm} This\hspace{-0.05cm} is\hspace{-0.05cm} also\hspace{-0.05cm} referred\hspace{-0.05cm} to\hspace{-0.05cm} as\hspace{-0.05cm} stream control \cite{r14}, i.e.,\hspace{-0.05cm} the\hspace{-0.05cm} number\hspace{-0.05cm} of independent input\hspace{-0.05cm} data streams for\hspace{-0.05cm} each\hspace{-0.05cm} user\hspace{-0.05cm} is\hspace{-0.05cm} limited.\hspace{-0.05cm} It\hspace{-0.05cm} can\hspace{-0.05cm} be\hspace{-0.05cm} interpreted as selectively shutting down some transmission antennas, which creates more interference to other users compared to the desired signal power generated for their target receivers.
\item\textit{Type II:} The transmit covariance matrices $\mathbf{Q}_i$ have different unitary matrices $\mathbf{V}_i$, but the same diagonal matrices $\mathbf{\Lambda}_i$ as $\mathbf{Q}_i^*$. This corresponds to the power control when all the effective users\hspace{-0.05cm} are\hspace{-0.05cm} the\hspace{-0.05cm} same,\hspace{-0.05cm} but\hspace{-0.05cm} their\hspace{-0.05cm} emitting\hspace{-0.05cm} directions\hspace{-0.05cm} are\hspace{-0.05cm} different\hspace{-0.05cm} from\hspace{-0.05cm} those\hspace{-0.05cm} of\hspace{-0.05cm} the\hspace{-0.05cm} NE.
\item\textit{Type III:} Both the unitary matrices $\mathbf{V}_i$ and diagonal matrices $\mathbf{\Lambda}_i$ of the transmit covariance matrices $\mathbf{Q}_i$ for the NB are different from those of the NE. This can be viewed as a combination of Type I and II of the optimal transmit covariance matrices for the NB.
\end{itemize}

\subsection{Interference Cancellation in 2-User Systems} \label{InterfCanc}

Based on Remark~1, if the interference is high, interference
cancellation techniques outperform the orthogonal signaling
techniques for guaranteeing the convexity of the rate region in the sense that applying interference cancellation leads to a larger rate region and, correspondingly, admits an NB
solution with larger user rates than the orthogonal signaling based solution. Moreover, it ensures the uniqueness of NB solution.

As a special case, we propose a multi-stage interference
cancellation technique for 2-user systems.\footnote{Note that the
generalization to the multi-user case is straightforward, but it
is omitted here for brevity.} This technique will be used in the
 numerical studies section where the 2-user case is primarily
considered. It is worth noting that our objective is to demonstrate the advantages of the interference cancellation over orthogonal signaling for the multicriteria optimization via NB rather than considering practically appealing interference cancellation designs.

To design our interference cancellation technique, it is required
that the receivers of both users cooperate with each other and
perform multi-stage interference cancellation as follows. First,
the receivers of users~1~and~2 decode their signals of interest
and pass them to the receivers of users~2~and~1, respectively.
Using \eqref{eq1}, the decoded signals from receivers of
users~1~and~2 in the first stage of interference cancellation can
be written as:
\begin {eqnarray}\label{eq29}
\hat{\mathbf{x}}_1^{(1)} \!\!&=&\!\! \frac{1}{\sqrt{\rho_1}}
\mathbf{H}_{11}^{-1}\mathbf{y}_1 \nonumber \\
\hat{\mathbf{x}}_2^{(1)} \!\!&=&\!\! \frac{1}{\sqrt{\rho_2}}
\mathbf{H}_{22}^{-1} \mathbf{y}_2
\end {eqnarray}
where $\hat{\mathbf{x}}_i^{(1)}$ $(i=1,2)$ are the decoded signals
at the receivers of users~1~and~2 in the first stage of the
interference cancellation procedure. Second, after receiving the
decoded signals from another user, the receivers of users~1~and~2
reconstruct the interference signals from the decoded signals,
which can be expressed as $\sqrt{\eta_{12}} \mathbf{H}_{12}
\hat{\mathbf{x}}_2 ^{(1)}$ and $\sqrt{\eta_{21}} \mathbf{H}_{21}
\hat{\mathbf{x}}_1 ^{(1)}$ for users~1~and~2, respectively.
Finally, each receiver subtracts the reconstructed interfering
signal from its originally received signal so to obtain the
following signals used in the second stage of the interference
cancellation procedure
\begin {eqnarray}\label{eq30}
\mathbf{y}_{1}^{(2)} \!\!&=&\!\! \mathbf{y}_{1} - \sqrt{\eta_{12}}
\mathbf{H}_{12}\hat{\mathbf{x}}_2^{(1)} \nonumber \\
\!\!&=&\!\! \left( \sqrt{\rho_1} {\mathbf{H}}_{11} - \sqrt{
\frac{\eta_{12} \eta_{21}}{\rho_{2}}} \mathbf{H}_{12}
\mathbf{H}_{22}^{-1} \mathbf{H}_{21}\right) {\mathbf{x}}_1 +
\mathbf{n}_1 - \sqrt{\frac{\eta_{12}}{\rho_{2}}} \mathbf{H}_{12}
\mathbf{H}_{22}^{-1} \mathbf{n}_2 \\
\mathbf{y}_{2}^{(2)} \!\!&=&\!\! \mathbf{y}_{2} - \sqrt{\eta_{21}}
\mathbf{H}_{21}\hat{\mathbf{x}}_1^{(1)} \nonumber \\
\!\!&=&\!\! \left(\sqrt{\rho_2} {\mathbf{H}}_{22} - \sqrt{
\frac{\eta_{21} \eta_{12}}{\rho_{1}}} \mathbf{H}_{21}
\mathbf{H}_{11}^{-1} \mathbf{H}_{12} \right) {\mathbf{x}}_2 +
\mathbf{n}_2 - \sqrt{ \frac{\eta_{21}} {\rho_{1}}}\mathbf{H}_{21}
\mathbf{H}_{11}^{-1} \mathbf{n}_1 \label{1stage}
\end {eqnarray}
where $\mathbf{y}_{i}^{(2)}\ (i=1,2)$ are the input signals at the
second stage of interference cancellation. The procedure is
repeated in the second stage to obtain the signals
\begin {eqnarray}\label{eq31}
\mathbf{y}_{1}^{(3)} \!\!\!&=&\!\!\! \mathbf{y}_{1} -
\sqrt{\eta_{12}} \mathbf{H}_{12} \left( \sqrt{\rho_2}
{\mathbf{H}}_{22} - \sqrt{ \frac{\eta_{21} \eta_{12}}
{\rho_{1}}} \mathbf{H}_{21} \mathbf{H}_{11}^{-1}
\mathbf{H}_{12} \right)^{-1} \mathbf{y}_2^{(2)} \nonumber \\
\!\!\!&=&\!\!\! \sqrt{\rho_1} \mathbf{H}_{11} {\mathbf{x}}_1
\!+\! \mathbf{n}_1 \!-\! \sqrt{\eta_{12}} \mathbf{H}_{12}
\!\left(\!\sqrt{\rho_2} {\mathbf{H}}_{22} - \sqrt{
\frac{\eta_{21} \eta_{12}}{\rho_{1}}} \mathbf{H}_{21}
\mathbf{H}_{11}^{-1} \mathbf{H}_{12}\!\right)^{-1} \!
\nonumber \\
\!\!& & \qquad \times \left(\! \mathbf{n}_2 \!-\!
\sqrt{\frac{\eta_{21}} {\rho_{1}}}
\mathbf{H}_{21}\mathbf{H}_{11}^{-1} \mathbf{n}_1\!\right) \\
\mathbf{y}_{2}^{(3)} \!\!\!&=&\!\!\! \mathbf{y}_{2} -
\sqrt{\eta_{21}} \mathbf{H}_{21} \left( \sqrt{\rho_1}
{\mathbf{H}}_{11} - \sqrt{\frac{\eta_{12} \eta_{21}} {\rho_{2}}}
\mathbf{H}_{12} \mathbf{H}_{22}^{-1} \mathbf{H}_{21}\right)^{-1}
\mathbf{y}_1^{(2)}
\nonumber \\
\!\!\!&=&\!\!\! \sqrt{\rho_{2}} \mathbf{H}_{22} {\mathbf{x}}_2
\!+\! \mathbf{n}_2 \!-\! \sqrt{\eta_{21}} \mathbf{H}_{21}
\!\left(\!\sqrt{\rho_1} {\mathbf{H}}_{11} \!-\! \sqrt{
\frac{\eta_{12}\eta_{21}} {\rho_{2}}} \mathbf{H}_{12}
\mathbf{H}_{22}^{-1} \mathbf{H}_{21} \right)^{-1} \nonumber \\
\!\!& & \qquad \times \left(\! \mathbf{n}_1 \!-\! \sqrt{
\frac{\eta_{12}}{\rho_{2}}} \mathbf{H}_{12} \mathbf{H}_{22}^{-1}
\mathbf{n}_2 \right) \label{2stage}
\end {eqnarray}
which are the input signals in the third stage of the procedure.
The decoding, reconstruction, and subtraction are terminated after
a certain stage or after a stopping criteria on the value of the
remaining error is satisfied. Note that it is guaranteed that at
each stage the new estimate of a signal is at least not worse than
the estimate in the previous stage \cite{multistage}.

\section{Numerical Studies}

\subsection{Convexity of the rate region}
Proposition 1 in Section III provides the sufficient condition guaranteeing the convexity of the rate region for MIMO interference systems. Unlike Proposition 1 where the interference-plus-noise covariance matrices and then INRs are interpreted in a qualitative manner, in this subsection we investigate the impact of the INRs on the convexity of the rate region quantitively via numerical studies. Consider a 2-user Rayleigh fading MIMO interference system with parameters $\sigma=1$ and $N_t=N_r=2$. We examine the probability that the rate region is convex for different values of SNR/INR. It can be seen in Fig.~1 that when the SNR is 0.1, 5 and 50, the SNR/INR should be at least 10, 15 and 20dB, respectively, to ensure that the rate region is convex. Two trends can be observed in Fig.~1: (i) as the SNR/INR increases, i.e., as the INR becomes decreasingly small while the SNR is fixed, the probability that the rate region is convex increases; (ii) as the SNR increases, the SNR/INR needs to increase as well to retain the probability that the rate region is convex.

When the Pareto  rate region is nonconvex, we can resort to either
mixed strategies, e.g., time-sharing or orthogonal signaling, or
interference cancellation techniques to construct a convex rate
region for the NB. Fig. 2 gives an example of these convexifying
schemes and their resulting convex rate regions for a 2-user MIMO
interference system with $N_t=N_r=2$, $\mathbf{H}_{11}= \mathrm{diag}(1.8,
1)$, $\mathbf{H}_{22}= \mathrm{diag}(1, 1.8)$, $\mathbf{H}_{12}$ =
$\mathbf{H}_{21}= \mathrm{diag}(1, 1)$, $\rho_1=\rho_2=20$ ,
$\eta_{12}=\eta_{21}=14$ and $p_1=p_2=1$.

Fig.~2 shows the maximum achievable rates for both users in the
cases when (i)~pure strategies are used, i.e., neither
time-sharing nor orthogonal signaling is adopted\footnote{For the
sake of brevity, we name the resulting rate region/boundary as the
pure-strategy rate region/boundary in the sequel.},
(ii)~time-sharing is employed, (iii)~orthogonal signaling is
employed, i.e., FDM and TDM are used, (iv)~the multi-stage
interference cancellation proposed in Subsection~\ref{InterfCanc}
is applied. Specifically, a 2-stage interference cancellation scheme
given in \eqref{eq31} and \eqref{2stage} is
considered. Its resulting rate boundary and the new NB solution is named as the IC boundary and NB solution after IC, respectively, in
the figure. The NB solution which can be obtained graphically from
the Nash curve\footnote{If the rates for users 1 and 2 at NE are
denoted as $I^{NE}_1$ and $I^{NE}_2$, the Nash curve can be
algebraically expressed as
$(y-I^{NE}_2)(x-I^{NE}_1)=\mathrm{max}(\mbox{Nash\ product})$. The NB
solution can be then interpreted as the intersection of the rate
boundary and the Nash curve as shown in Fig. 2.} and the NE for
this MIMO system are also shown in Fig.~2. It can be seen from the
figure that the orthogonal signaling schemes, especially the TDM,
result in significant rate loss. The time-sharing signaling has
a larger rate region than TDM and FDM. While, the
multi-stage interference cancellation leads to the largest rate
region. This example shows the potentials and advantages of interference
cancellation techniques over the orthogonal signaling
and time-sharing schemes in terms of guaranteeing a larger convex rate
region and, correspondingly, an NB solution with larger user rates. Moreover, the interference cancellation-based approach is
more desirable than the time-sharing signaling due to the fact
that the corresponding rates are actually achievable, while the
rates promised by the time-sharing-based approach are achievable
only in average. Note that practically instantaneous rates are more
important than average rates and so the desired instantaneous rate must be~achievable.

\subsection{Fairness of the NB}

NB is a bargaining approach which balances the individual fairness and system-level efficiency. The fairness of the NB is guaranteed by the fact that its resultant utility for each user is not less than that of the NE, which is considered as a relatively fair approach due to its selfish and competitive nature.  In this subsection, we study the fairness of the NB solution and compare it with other bargaining approaches. The following bargaining solutions from \cite{solutions} are taken as the benchmark for comparison:
\begin{itemize}
\item Egalitarian solution: It is an absolutely fair solution with identical rate for each user.
\item Kalai-Smorodinsky (K-S) solution: It results in utilities proportional to their maximal achievable rates.
\item Utilitarian solution: This solution maximizes the sum rate of all users.
\item Proportional solution: It maximizes the rate product of all users.
\end{itemize}
We compare the fairness of these bargaining solutions with the NB solution in terms of Jain's fairness index (JFI)\cite{jain} having the Egalitarian solution as the optimal solution. Recently, the JFI has been also used to compare difference scheduling algorithms in MIMO broadcast channels~\cite{jfi_mimo}. JFI $J$ for an $n$-user system is obtained by Jain's equation\cite{jain}
\begin {equation}\label{eq32}
J = \frac{(\sum_{i=1}^{n}x_i)^2}{n \sum_{i=1}^{n}x_i^2}
\end {equation}
where $x_i=T_i/T_i^*$, $\{T_1,\ T_2, \cdots, T_n\}$ and $\{T_1^*,\ T_2^*, \cdots, T_n^*\}$ are the measured and optimal utility vectors for the $n$-user system, respectively. It rates the fairness of the system and ranges from $1/n$ (worst case) to 1 (best case).

To compare the fairness of the NB solution with that of the above mentioned bargaining solutions, we consider a 2-user MIMO interference system as an example with $N_t=N_r=2$, $\mathbf{H}_{11} = \mathrm{diag}(1.8, 1)$, $\mathbf{H}_{22} = \mathrm{diag}(0.5, 0.6)$, $\mathbf{H}_{12} = \mathrm{diag}(1, 1)$, $\mathbf{H}_{21} = \mathrm{diag}(0.3, 0.3)$, $\rho_1=\rho_2=20$ , $\eta_{12}=\eta_{21}=14$ and $p_1=p_2=1$. Fig.~3 shows the pure-strategy rate boundary, time-sharing rate boundary and different bargaining solutions for this MIMO interference system. These bargaining solutions are demonstrated graphically and geometrically in Fig.~3. For example, the NB solution can be interpreted as the intersection of the rate boundary and the Nash curve. Similarily, the Egalitarian solution, K-S solution, Utilitarian solution and Proportional solution are the intersections of the rate boundary and the following curves $y=x$, $y/x=\mathrm{max}(I_2)/\mathrm{max}(I_1)$, $y+x=\mathrm{max}(I_1+I_2)$, $yx=\mathrm{max}(I_1\cdot I_2)$, respectively. JFIs obtained from~\eqref{eq32} for the NE, NB solution, K-S solution, Utilitarian solution and Proportional solution are 0.9925, 0.9685, 0.8853, 0.8960 and 0.8960, respectively. Note that in this example, the Utilitarian solution is identical to the Proportional solution. It can be seen that the JFI of the NB solution is only slightly smaller than that of the NE but much larger than all the other solutions. Compared with the NE, the NB improves the sum rate of this MIMO system by $18\%$ at the price of only compromising its JFI by 0.9925-0.9685=0.024. These results confirm that the NB is an effective approach well balancing the indivisual fairness and system performance of interference systems.

\subsection{Feasibility of NB Set}

To exemplify the feasibility of NB set discussed in Section III-B, we first consider a 2-user MIMO interference system with the same setup as Fig.~2. In this example, the transmit covariance matrices of the NE obtained by the IWF are $\mathbf{Q}_1^{N\hspace{-0.06cm} E} = \mathrm{diag}(0.25, 0.75)$ and $\mathbf{Q}_2^{N\hspace{-0.06cm} E} = \mathrm{diag}(0.75, 0.25)$. The transmit covariance matrices resulting in the NB are $\mathbf{Q}_1^{N\hspace{-0.06cm} B}= \mathrm{diag}(1,0)$ and $\mathbf{Q}_2^{N\hspace{-0.06cm} B} = \mathrm{diag}(0,1)$. As we can see, the transmit covariance matrices $\mathbf{Q}_i^{N\hspace{-0.06cm} B}$ $(i=1,2)$ are of Type I as mentioned in the previous section, with the same unitary matrices but different diagonal matrices from $\mathbf{Q}_i^{N\hspace{-0.06cm} E}$.

As a second example, consider a 2-user MIMO interference system with
\begin {equation*}
N_t=N_r=2,\ \mathbf{H}_{11}=\mathbf{H}_{22}=
\left(\begin{array}{cc}
\hspace{-0.15cm} 2\hspace{-0.2cm}&2 \hspace{-0.15cm}\\  \hspace{-0.15cm}1\hspace{-0.2cm}&1 \hspace{-0.15cm}
\end{array}\right)
,\
\mathbf{H}_{12}=\mathbf{H}_{21}=
\left(\begin{array}{cc}
\hspace{-0.15cm} 1\hspace{-0.2cm}&0 \hspace{-0.15cm}\\  \hspace{-0.15cm}0\hspace{-0.2cm}&1 \hspace{-0.15cm}
\end{array}\right)
,
\end {equation*}
$\rho_1\hspace{-0.05cm}=\hspace{-0.05cm}\rho_2\hspace{-0.05cm}=\hspace{-0.05cm}20$, $\eta_{12}\hspace{-0.05cm}=\hspace{-0.05cm}\eta_{21}\hspace{-0.05cm}=\hspace{-0.05cm}60$ and $p_1\hspace{-0.05cm}=\hspace{-0.05cm}p_2\hspace{-0.05cm}=\hspace{-0.05cm}1$.  Fig.~4 demonstrates the pure-strategy rate boundary, the NE and NB solutions. It can also be seen from Fig. 4 that the rate region is convex, so there is no need to employ signaling. The transmit covariance matrices of the NE in this example are
\begin {equation*}
\mathbf{Q}_1^{N\hspace{-0.06cm} E}=\mathbf{Q}_2^{N\hspace{-0.06cm} E}=
\left(\begin{array}{cc}
\hspace{-0.15cm} 0.50\hspace{-0.2cm}&0.50 \hspace{-0.15cm}\\  \hspace{-0.15cm}0.50\hspace{-0.2cm}&0.50 \hspace{-0.15cm}
\end{array}\right)
=
\left(\begin{array}{rr}
\hspace{-0.15cm} -0.71\hspace{-0.2cm}&0.71 \hspace{-0.15cm}\\  \hspace{-0.15cm}0.71\hspace{-0.2cm}&0.71 \hspace{-0.15cm}
\end{array}\right)
\left(\begin{array}{cc}
\hspace{-0.15cm} 0\hspace{-0.2cm}&0 \hspace{-0.15cm}\\  \hspace{-0.15cm}0\hspace{-0.2cm}&1 \hspace{-0.15cm}
\end{array}\right)
\left(\begin{array}{rr}
\hspace{-0.15cm} -0.71\hspace{-0.2cm}&0.71 \hspace{-0.15cm}\\  \hspace{-0.15cm}0.71\hspace{-0.2cm}&0.71 \hspace{-0.15cm}
\end{array}\right)
.
\end {equation*}
Whereas, the transmit covariance matrices leading to the NB solution are
\begin {equation*}
\mathbf{Q}_1^{N\hspace{-0.06cm} B}=\mathbf{Q}_2^{N\hspace{-0.06cm}B}\hspace{-0.05cm}=\hspace{-0.05cm}
\left(\begin{array}{rr}
\hspace{-0.15cm} 0.02\hspace{-0.2cm}&0.15 \hspace{-0.15cm}\\  \hspace{-0.15cm}0.15\hspace{-0.2cm}&0.98 \hspace{-0.15cm}
\end{array}\right)
=
\left(\begin{array}{rr}
\hspace{-0.15cm} -0.99\hspace{-0.2cm}&0.16 \hspace{-0.15cm}\\  \hspace{-0.15cm}0.16\hspace{-0.2cm}&0.99 \hspace{-0.15cm}
\end{array}\right)
\left(\begin{array}{rr}
\hspace{-0.15cm} 0\hspace{-0.2cm}&0 \hspace{-0.15cm}\\  \hspace{-0.15cm}0\hspace{-0.2cm}&1 \hspace{-0.15cm}
\end{array}\right)
\left(\begin{array}{rr}
\hspace{-0.15cm} -0.99\hspace{-0.2cm}&0.16 \hspace{-0.15cm}\\  \hspace{-0.15cm}0.16\hspace{-0.2cm}&0.99 \hspace{-0.15cm}
\end{array}\right).
\end {equation*}
In this scenario, the transmit covariance matrices $\mathbf{Q}_i^{N\hspace{-0.06cm} B}$ have the structure of Type II, i.e., they have the same diagonal matrices but different unitary matrices. Alternatively, the transmit covariance matrix can be interpreted as precoding. The corresponding precoding matrices for its NB and NE are
\begin {equation*}
\mathbf{F}_i^{N\hspace{-0.06cm} B}=
\left(\begin{array}{rr}
\hspace{-0.15cm} 0\hspace{-0.15cm}&0.16 \hspace{-0.15cm}\\  \hspace{-0.15cm}0\hspace{-0.15cm}&0.99 \hspace{-0.15cm}
\end{array}\right)
,\
\mathbf{F}_i^{N\hspace{-0.06cm} E}=
\left(\begin{array}{rr}
\hspace{-0.15cm} 0\hspace{-0.15cm}&0.71 \hspace{-0.15cm}\\  \hspace{-0.15cm}0\hspace{-0.15cm}&0.71 \hspace{-0.15cm}
\end{array}\right)
.
\end {equation*}

As the last example, consider another 2-user MIMO interference system with
\begin {equation*}
N_t=N_r=2,\ \mathbf{H}_{11}=
\left(\begin{array}{cc}
\hspace{-0.15cm} 2\hspace{-0.2cm}&1 \hspace{-0.15cm}\\  \hspace{-0.15cm}2\hspace{-0.2cm}&1 \hspace{-0.15cm}
\end{array}\right)
,\
\mathbf{H}_{22}=
\left(\begin{array}{cc}
\hspace{-0.15cm} 1\hspace{-0.2cm}&0 \hspace{-0.15cm}\\  \hspace{-0.15cm}0\hspace{-0.2cm}&2 \hspace{-0.15cm}
\end{array}\right)
,\
\mathbf{H}_{12}=\mathbf{H}_{21}=
\left(\begin{array}{cc}
\hspace{-0.15cm} 1\hspace{-0.2cm}&0 \hspace{-0.15cm}\\  \hspace{-0.15cm}0\hspace{-0.2cm}&1 \hspace{-0.15cm}
\end{array}\right)
,
\end {equation*}
$\rho_1\hspace{-0.05cm}=\hspace{-0.05cm}\rho_2\hspace{-0.05cm}=\hspace{-0.05cm}20$, $\eta_{12}\hspace{-0.05cm}=\hspace{-0.05cm}\eta_{21}\hspace{-0.05cm}=\hspace{-0.05cm}16$ and $p_1\hspace{-0.05cm}=\hspace{-0.05cm}p_2\hspace{-0.05cm}=\hspace{-0.05cm}1$. Fig.~5 shows the time-sharing rate boundary, pure-strategy rate boundary, the NE and NB solutions. It can be seen from Fig.~5 that the rate region is nonconvex in this MIMO system, but its NB solution still lies on the pure-strategy rate boundary. The transmit covariance matrices of the NE in this MIMO interference system~are
\vspace{-0.5cm}
\begin {equation*}
\mathbf{Q}_1^{N\hspace{-0.06cm} E}=
\left(\begin{array}{cc}
\hspace{-0.15cm} 0.80\hspace{-0.2cm}&0.40 \hspace{-0.15cm}\\  \hspace{-0.15cm}0.40\hspace{-0.2cm}&0.20 \hspace{-0.15cm}
\end{array}\right)
=
\left(\begin{array}{rr}
\hspace{-0.15cm} -0.89\hspace{-0.2cm}&-0.45 \hspace{-0.15cm}\\  \hspace{-0.15cm}-0.45\hspace{-0.2cm}&0.89 \hspace{-0.15cm}
\end{array}\right)
\left(\begin{array}{cc}
\hspace{-0.15cm} 1\hspace{-0.2cm}&0 \hspace{-0.15cm}\\  \hspace{-0.15cm}0\hspace{-0.2cm}&0 \hspace{-0.15cm}
\end{array}\right)
\left(\begin{array}{rr}
\hspace{-0.15cm} -0.89\hspace{-0.2cm}&-0.45 \hspace{-0.15cm}\\  \hspace{-0.15cm}-0.45\hspace{-0.2cm}&0.89 \hspace{-0.15cm}
\end{array}\right)
\end {equation*}

\begin {equation*}
\mathbf{Q}_2^{N\hspace{-0.06cm} E}=
\left(\begin{array}{cc}
\hspace{-0.15cm} 0.22\hspace{-0.2cm}&-0.14 \hspace{-0.15cm}\\  \hspace{-0.15cm}-0.14\hspace{-0.2cm}&0.78 \hspace{-0.15cm}
\end{array}\right)
=
\left(\begin{array}{rr}
\hspace{-0.15cm} -0.23\hspace{-0.2cm}&0.97 \hspace{-0.15cm}\\  \hspace{-0.15cm}0.97\hspace{-0.2cm}&0.23 \hspace{-0.15cm}
\end{array}\right)
\left(\begin{array}{cc}
\hspace{-0.15cm} 0.81\hspace{-0.2cm}&0 \hspace{-0.15cm}\\  \hspace{-0.15cm}0\hspace{-0.2cm}&0.19 \hspace{-0.15cm}
\end{array}\right)
\left(\begin{array}{rr}
\hspace{-0.15cm} -0.23\hspace{-0.2cm}&0.97 \hspace{-0.15cm}\\  \hspace{-0.15cm}0.97\hspace{-0.2cm}&0.23 \hspace{-0.15cm}
\end{array}\right)
,
\end {equation*}
and the transmit covariance matrices leading to the NB solution are
\begin {equation*}
\mathbf{Q}_1^{N\hspace{-0.06cm} B}=\hspace{-0.05cm}
\left(\begin{array}{rr}
\hspace{-0.15cm} 0.90\hspace{-0.2cm}&0.29 \hspace{-0.15cm}\\  \hspace{-0.15cm}0.29\hspace{-0.2cm}&0.10 \hspace{-0.15cm}
\end{array}\right)
=
\left(\begin{array}{rr}
\hspace{-0.15cm} -0.95\hspace{-0.2cm}&-0.31 \hspace{-0.15cm}\\  \hspace{-0.15cm}-0.31\hspace{-0.2cm}&0.95 \hspace{-0.15cm}
\end{array}\right)
\left(\begin{array}{rr}
\hspace{-0.15cm} 1\hspace{-0.2cm}&0 \hspace{-0.15cm}\\  \hspace{-0.15cm}0\hspace{-0.2cm}&0 \hspace{-0.15cm}
\end{array}\right)
\left(\begin{array}{rr}
\hspace{-0.15cm} -0.95\hspace{-0.2cm}&-0.31 \hspace{-0.15cm}\\  \hspace{-0.15cm}-0.31\hspace{-0.2cm}&0.95 \hspace{-0.15cm}
\end{array}\right)
\end {equation*}
\begin {equation*}
\mathbf{Q}_2^{N\hspace{-0.06cm} B}=\hspace{-0.05cm}
\left(\begin{array}{rr}
\hspace{-0.15cm} 0.21\hspace{-0.2cm}&-0.40 \hspace{-0.15cm}\\  \hspace{-0.15cm}-0.40\hspace{-0.2cm}&0.79 \hspace{-0.15cm}
\end{array}\right)
=
\left(\begin{array}{rr}
\hspace{-0.15cm} -0.45\hspace{-0.2cm}&0.89 \hspace{-0.15cm}\\  \hspace{-0.15cm}0.89\hspace{-0.2cm}&0.45 \hspace{-0.15cm}
\end{array}\right)
\left(\begin{array}{rr}
\hspace{-0.15cm} 1\hspace{-0.2cm}&0 \hspace{-0.15cm}\\  \hspace{-0.15cm}0\hspace{-0.2cm}&0 \hspace{-0.15cm}
\end{array}\right)
\left(\begin{array}{rr}
\hspace{-0.15cm} -0.45\hspace{-0.2cm}&0.89 \hspace{-0.15cm}\\  \hspace{-0.15cm}0.89\hspace{-0.2cm}&0.45 \hspace{-0.15cm}
\end{array}\right).
\end {equation*}
As we can see, $\mathbf{Q}_1^{N\hspace{-0.06cm}B}$ and $\mathbf{Q}_2^{N\hspace{-0.06cm}B}$ have the structure of Types II and III, respectively, i.e., both the unitary and diagonal matrices for $\mathbf{Q}_2^{N\hspace{-0.06cm}B}$ are different from those of the corresponding NE.

\subsection{The SNR and INR Impact on the Existence of the NB Solution}

Figs. 6 and 7 depict the SNR and INR impact on the NE and NB solutions, respectively. The channel realization is the same as that used for obtaining Fig. 4. It can be seen from Fig.~6 that the NB solution provides better rates as the SNR increases. Another phenomenon is that at low and high SNRs the NE coincides with the NB solution, i.e., the NE is the optimal solution. This is due to the fact that the NE lies on the rate boundary at these SNRs. In this example, the NE is optimal when SNR is less than 15 or larger than 230. In Fig.~7, the NB solution remains unchanged in the INR range where the NB solution exists. This is due to the fact that the NB in this case adopts stream control, i.e., only one stream is transmitting for each user, which eventually converts the MIMO interference system into an interference-free system.

Fig. 8 presents the existence of the NB solution at different values of the SNR and INR. The symbol `$\times$' represents the fact that the NB solution exists at the corresponding SNR and INR, i.e., the rate of the users in NE can be further improved by simultaneously changing their transmit covariance matrices from the NE. As mentioned in Section III-B, for systems with different channel matrices, the SNR and INR impact on the existence of their NB solutions may vary. In this MIMO interference system, the NB solution exists only when the INR is smaller than the SNR. From Figs.~7 and 8, it can also be seen that the NB solution does not exist when the INR is too small. The reasons behind this phenomenon are as follows:
\begin {itemize}
\item When the INR is very small compared to the SNR, if stream control is adopted, the negative contribution to the rate from the loss in the desired signal component will be larger than the positive contribution from the decrease in interference. It is preferable for both users to compete selfishly with each other rather than adopting stream control. Therefore, in this very small INR region which is depicted as the selfish competition region in Fig. 8, the NE and NB solutions are identical.
\item As the INR increases and exceeds a certain value, the negative contribution to the rate from the loss in desired signal component becomes smaller than the positive contribution from the decrease in interference after adopting stream control, but the interference is still not strong enough to lead the IWF to converge to the stream control solution. In this case, the NB ends up with the stream control solution but the NE does not, and the NB solution exists.
\item As the INR continues increasing, the interference becomes stronger, which stimulates the IWF to approach the stream control solution. When the INR reaches or exceeds a certain larger value, the NE converges to the stream control solution. In this region with large INR (we name it as the stream control region in Fig. 8), the NE coincides with the NB solution again.
\end {itemize}

Another phenomenon can be observed from Figs. 6 and 8 is that the NB solution does not exist when the SNR is relatively small or large. This phenomenon can be explained similarly as the previous one:
\begin {itemize}
\item When the SNR is small compared to the INR, the IWF is driven to the stream control solution which is also the optimal solution in that region, i.e., the NE and NB solutions are the same, both of them choose stream control.
\item As the SNR increases and exceeds a certain threshold, the IWF is driven to deviate from stream control and approach selfish competition, but the optimal (NB) solution in this area is still stream control. In this region, the NB solution exists and outperforms the NE solution which adopts selfish competition.
\item As the SNR continues increasing, the NB solution begins to approach selfish competition since the interference becomes less significant compared to the desired signal. When the SNR reaches a certain larger value, the NB solution converges to selfish competition which is same as the NE solution.
\end {itemize}
\subsection{Uniqueness of the NB Solution}

As a natural extension to our numerical studies of the existence of the NB solution, we also examine the uniqueness of the NB solution by calculating the value of $f^{''}(t)$ in \eqref{eq21} when the NB solution exists. It can be seen from \eqref{eq21}--\eqref{eq24} that $f^{''}(t)$ is a function of the underlying channel (channel matrices, SNRs and INRs). Meanwhile, it is also a function of $t$, $(\mathbf{Z}_1,\mathbf{Z}_2)$ and $(\mathbf{X}_1,\mathbf{X}_2)$ in the convex combination
\eqref{eq10}. Let the channel matrices be the same as those used to obtain Fig. 2. We set $t=0.5$, and choose transmit covariance matrices $(\mathbf{Z}_1,\mathbf{Z}_2)$ and $(\mathbf{X}_1,\mathbf{X}_2)$ such that their corresponding rates would be in the maximum achievable rate boundary, their rates for user 1 are $I_1^{N\hspace{-0.06cm} B}$ and $I_1^{N\hspace{-0.06cm} B}$/2, respectively. With this setup, Fig.~9 depicts the value of $f^{''}(t)|_{t=0.5}$. It can be seen that in this instance the value of $f^{''}(t)$ is negative in the region where the NB exists. We can infer from Section III-B that when the NB solution exists and the value of $f^{''}(t)$ is not positive for any $t$, $(\mathbf{Z}_1,\mathbf{Z}_2)$ and $(\mathbf{X}_1,\mathbf{X}_2)$, then its NB solution is unique.


\section{Special Cases}
In the previous section, we were concerned with characterizing the Pareto rate region and the NB over MIMO interference systems. Here, we further emphasize that some findings and results regarding the convexity of the rate region, the uniqueness, existence and the feasibility of the NB set have much broader validity than just in the MIMO case. Particularly, these results also hold for MIMO and SISO interference systems. In fact, the NB over MIMO interference systems unifies and generalizes that of  MISO and SISO scenarios. In what follows, the applicability of the Pareto rate region and some NB characteristics is analyzed in the case of MISO and SISO.

$\textbf{Remark\ 2}$: {\it The sufficient conditions which guarantee the convexity of the rate region and the uniqueness of the NB solution for MIMO interference systems derived in Section~III still hold in both MISO and SISO interference systems.}

The expression for the mutual information \eqref{eq2} is a general one and is applicable for MIMO, MISO and SISO systems. The channel matrix $\mathbf{H}_{ij}$ in \eqref{eq2} converts from an $N_r \times N_t$ matrix for MIMO to a $1 \times N_t$ vector $\mathbf{h}_{ij}$ and a complex value $h_{ij}$ for MISO and SISO, respectively. The transmit covariance matrix $\mathbf{Q}_{i}$ is an $N_t \times N_t$ matrix for both MIMO and MISO, and a non-negative real value $Q_i$ for SISO. More specifically, equation \eqref{eq2} can be written as
\begin{equation} \label{eq35}
I_i(\mathbf{Q}) = \left\{
\begin{array}{cc}
\log_2 \left(1 + \frac{\rho_i\mathbf{h}_{ii}\mathbf{Q}_i
{\mathbf{h}}_{ii}^H}{1+\sum_{j = 1, j\neq i}^{M} \eta_{ij}
{\mathbf{h}}_{ij} {\mathbf{Q}}_{j} {\mathbf{h}}_{ij}^H}
\right) & \text{for\ MISO}\\
\log_2 \left(1 + \frac{\rho_i{|{h}_{ii}|}^2{Q}_i}{1+
\sum_{j = 1, j\neq i}^{M} \eta_{ij} {|{h}_{ij}|}^2
{{Q}}_{j}} \right)  & \text{for\ SISO}\\
\end{array} \right. .
\end{equation}
The optimization problems \eqref{eq9} and \eqref{eq18} are still applicable to MISO and SISO systems. We further express the transmitted signal vector in \eqref{eq1} as $\mathbf{x}_i=\mathbf{F}_i\mathbf{s}_i$, where $\mathbf{F}_i$ is an $N_t \times N_r$ matrix which is known as precoding matrix in MIMO and $\mathbf{s}_i$ is the $N_r \times 1$ information symbol vector for user $i$. Without loss of generality, we assume that $E[{\mathbf{s}}_i{\mathbf{s}}_i^H]=\mathbf{I}$. Then, the transmit covariance matrix can be generalized as
\begin{equation}\label{eq36}
\mathbf{Q}_i = \left\{
\begin{array}{rl}
E[\mathbf{F}_i\mathbf{F}_i^H] \ &\text{for \ MIMO}\\
E[\mathbf{w}_i\mathbf{w}_i^H] \ &\text{for \ MISO}\\
E[|w_i|^2] \ &\text{for \ SISO}\\
\end{array} \right.
\end{equation}
where $\mathbf{w}_i$ is the $N_t \times 1$ beamforming vector for MISO transmitter $i$ and $w_i$ is the complex valued power control weight for SISO transmitter $i$.  Thus, the optimization problem \eqref{eq9} can be interpreted as follows: the NB aims at designing the optimal precoding matrices, beamforming vectors and power control weights for all cooperative users under the power constraints, with the objective to maximize the Nash product of the MIMO, MISO and SISO interference system, respectively. Therefore, sufficient conditions ensuring the convexity of the rate region and the uniqueness of the NB solution in MIMO interference systems derived in Section III-A and B can be extended to both MISO and SISO scenarios, i.e., Propostion 1$\sim$3 are still valid for MISO and SISO interference systems.

$\textbf{Remark\ 3}$: {\it The analytical structure of the transmit covariance matrices leading to the NB in Section~III-B and the numerical studies on existence of the NB solution in MIMO interference systems are still valid in MISO case.}

This is due to the fact that these two characteristics are analyzed based on the transmit covariance matrix $\mathbf{Q}_i$, which is an $N_t \times N_t$ Hermitian PSD matrix for both MIMO and MISO systems. However, as for a SISO interference system, the transmit covariance matrix $\mathbf{Q}_i$ degenerates to a non-negative real value $Q_i$, so these results are not applicable to SISO interference systems.


\section{Conclusions}

In this paper, the rate control problem in multi-user MIMO interference systems has been formulated as an MCO problem. The convexity of the Pareto rate region of this MCO problem has been studied. A sufficient condition which guarantees the convexity of the rate region has been derived. It is argued that the interference cancellation techniques are preferable for convexifying the Pareto rate region. Then, the MCO problem has been transformed into a single-objective optimization problem using NB. A variety of characteristics for the NB solution in MIMO interference systems such as the uniqueness and the existence of the NB solution and the feasibility of the NB set have been investigated. A sufficient condition ensuring the uniqueness of the NB solution in MIMO interference systems has also been derived. Moreover, a multi-stage interference cancellation scheme has been proposed to convexify the rate region of the corresponding MCO problem. It is shown that it leads to an NB solution with larger user rates. The convexity of the rate region, the effectiveness of the proposed interference cancellation scheme, the fairness of the NB solution, the impact of the SNR and INR on the NE and NB solutions, the existence of the NB solution, and its uniqueness (if exists) have also been demonstrated via numerical studies. Finally, the applicability of several NB characteristics to the MISO and SISO interference systems has been shown.

\vspace{1cm}
\begin{center}
\textsc{{Acknowledgments}}\\
\end{center}

Zengmao Chen, Cheng-Xiang Wang and John Thompson acknowledge the support from the Scottish Funding Council for the Joint Research Institute in Signal and Image Processing between the University of Edinburgh and Heriot-Watt University which is a part of the Edinburgh Research Partnership in Engineering and Mathematics (ERPem). Sergiy A. Vorobyov acknowledges the support in part from the Natural Sciences and Engineering Research Council (NSERC) of Canada and in part from the Alberta Ingenuity Foundation, Alberta, Canada.


\newpage


\begin{center}
\begin{tabular}{c}
\hskip-0.3cm\epsfxsize=14cm\epsffile{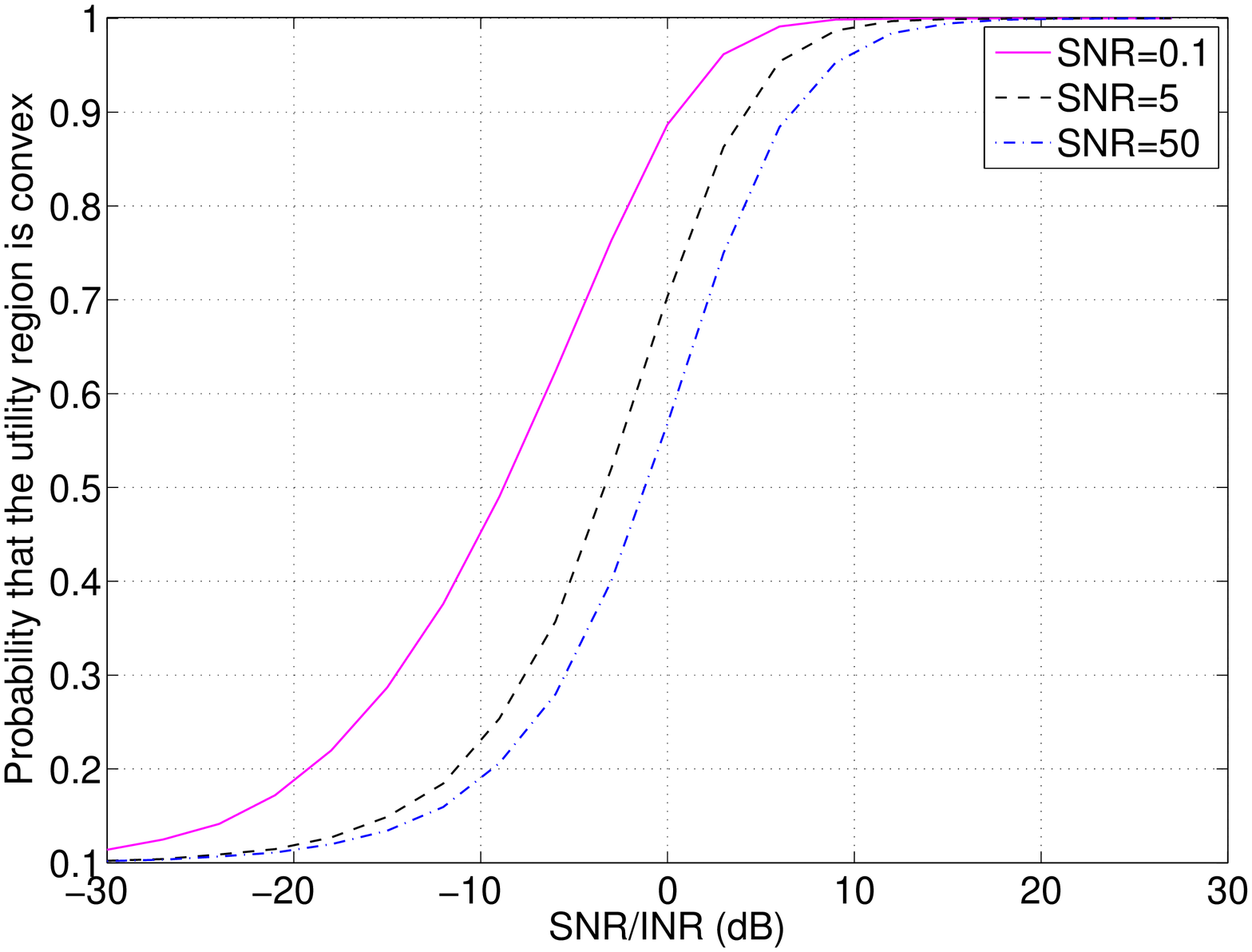} \\
\end{tabular}\\
\vspace*{-0.4cm}
\begin{center}
\small Fig.~1. Probability of the convexity of the rate region over different values of SNR/INR.
\end{center}
\end{center}

\begin{center}
\begin{tabular}{c}
\hskip-0.3cm\epsfxsize=14cm\epsffile{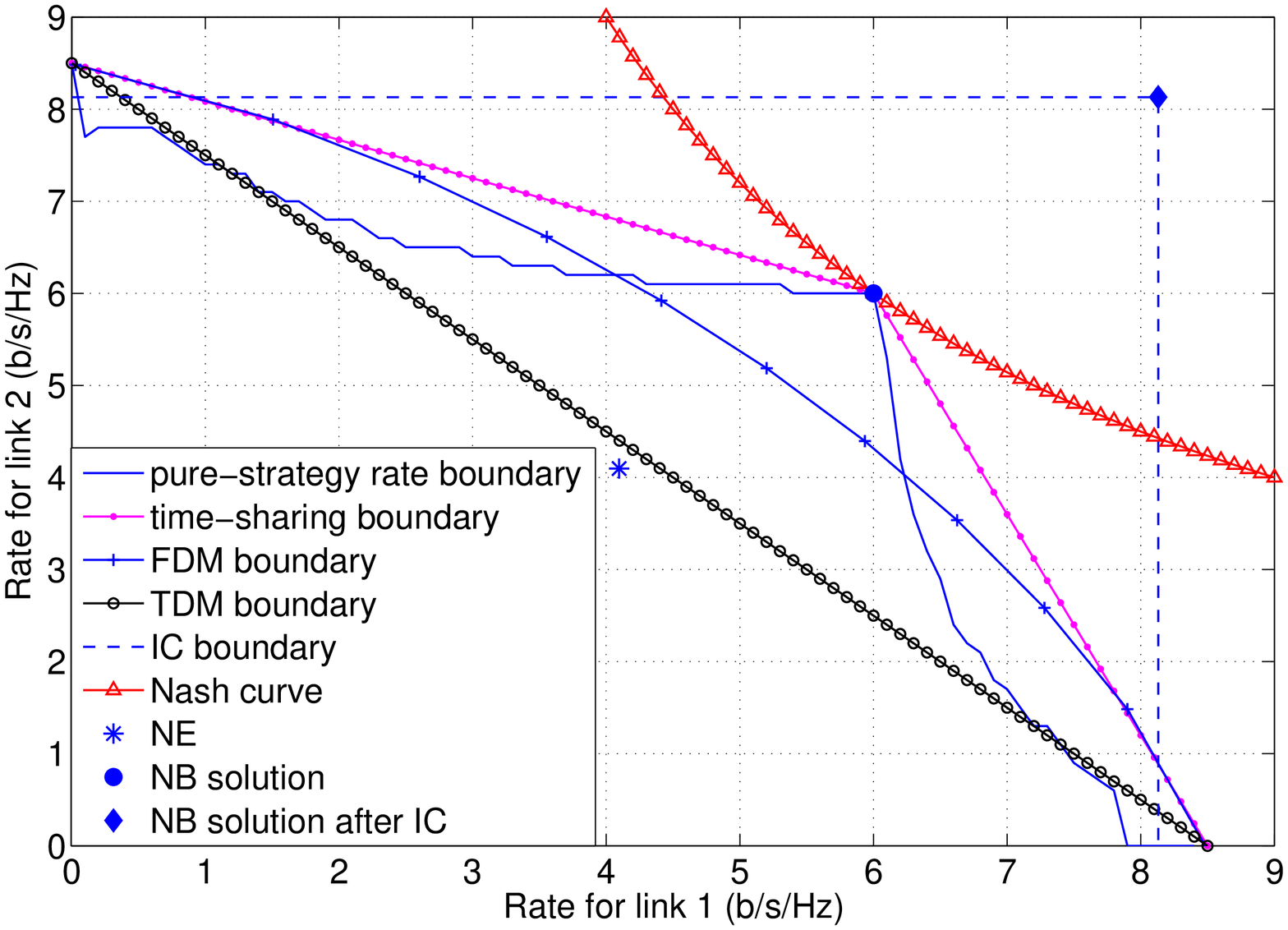} \\
\end{tabular}\\
\vspace*{-0.7cm}
\begin{center}
\small Fig.~2. An example of nonconvex rate region, different signaling schemes and NB solution with Type I feasible NB set.
\end{center}
\end{center}

\begin{center}
\begin{tabular}{c}
\hskip-0.3cm\epsfxsize=14cm\epsffile{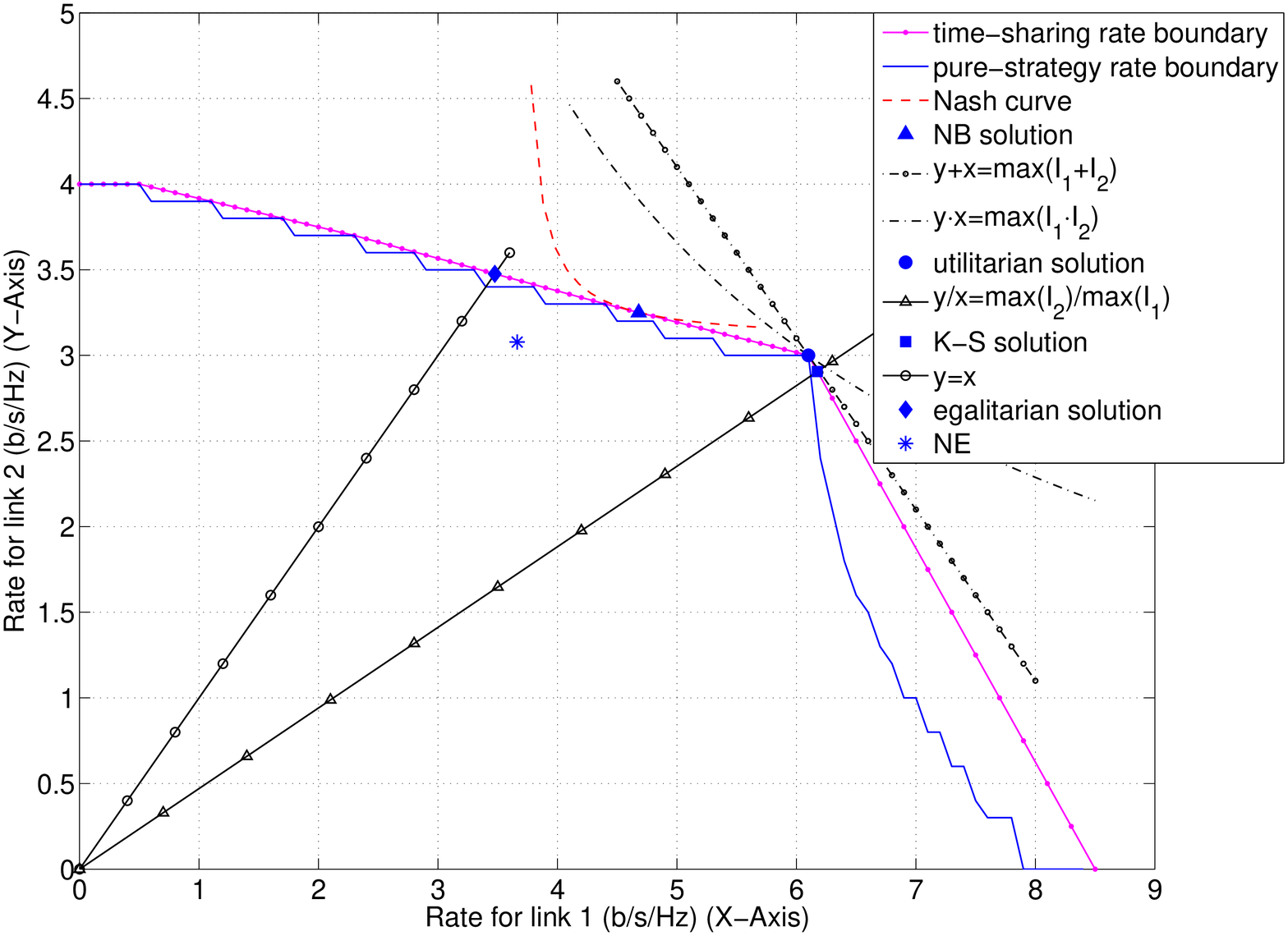} \\
\end{tabular}\\
\vspace*{-0.4cm}
\begin{center}
\small Fig.~3. Various bargaining solutions for a MIMO interference system.
\end{center}
\end{center}

\begin{center}
\begin{tabular}{c}
\hskip-0.3cm\epsfxsize=14cm\epsffile{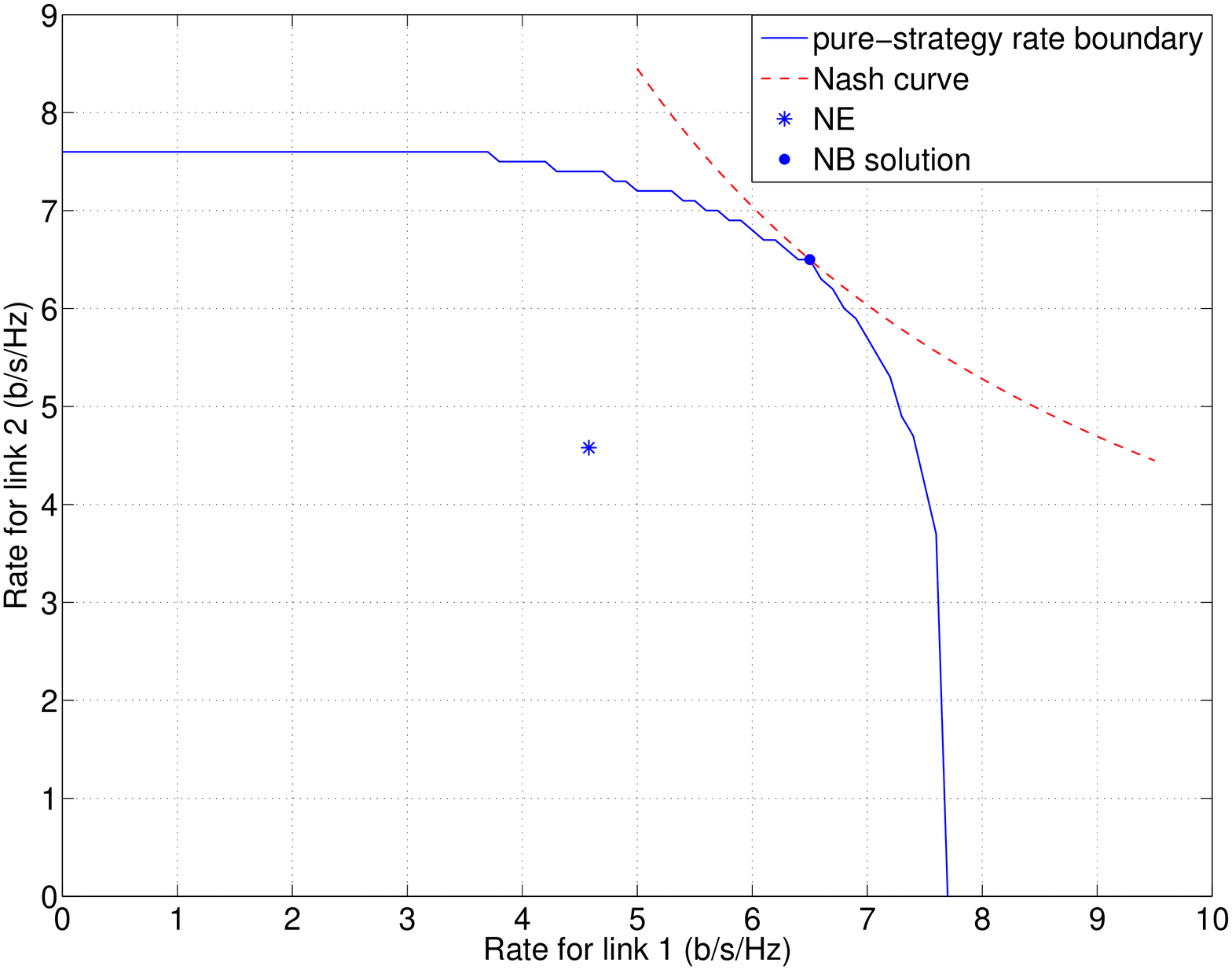} \\
\end{tabular}\\
\vspace*{-0.4cm}
\begin{center}
\small Fig.~4. Rate region, NE and NB solution with Type II feasible NB set.
\end{center}
\end{center}

\begin{center}
\begin{tabular}{c}
\hskip-0.6cm\epsfxsize=14cm\epsffile{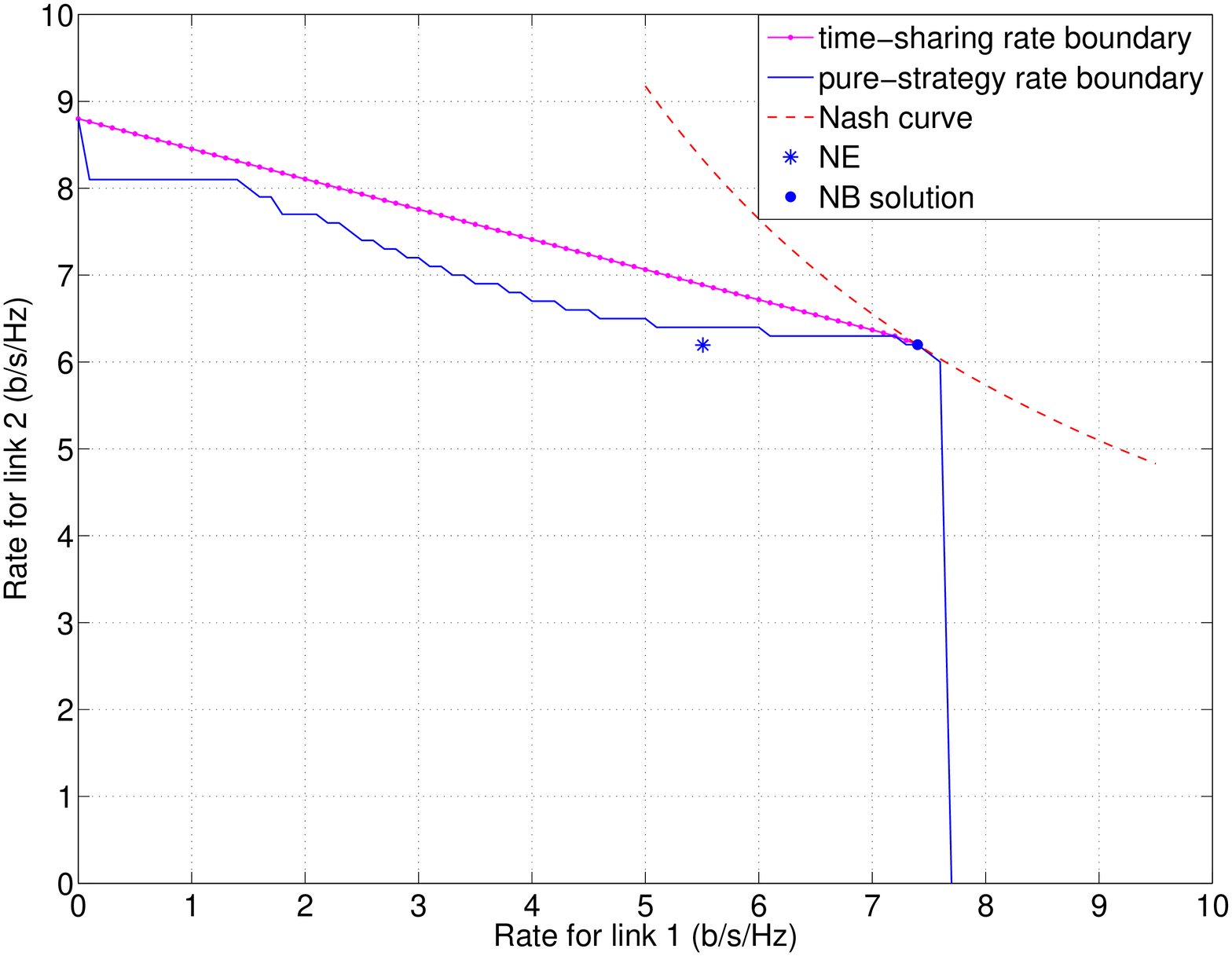} \\
\end{tabular}\\
\vspace*{-0.25cm}
\begin{center}
\small Fig.~5. Rate region, NE and NB solution with Type III feasible NB set.
\end{center}
\end{center}

\begin{center}
\begin{tabular}{c}
\hskip-0.6cm\epsfxsize=14cm\epsffile{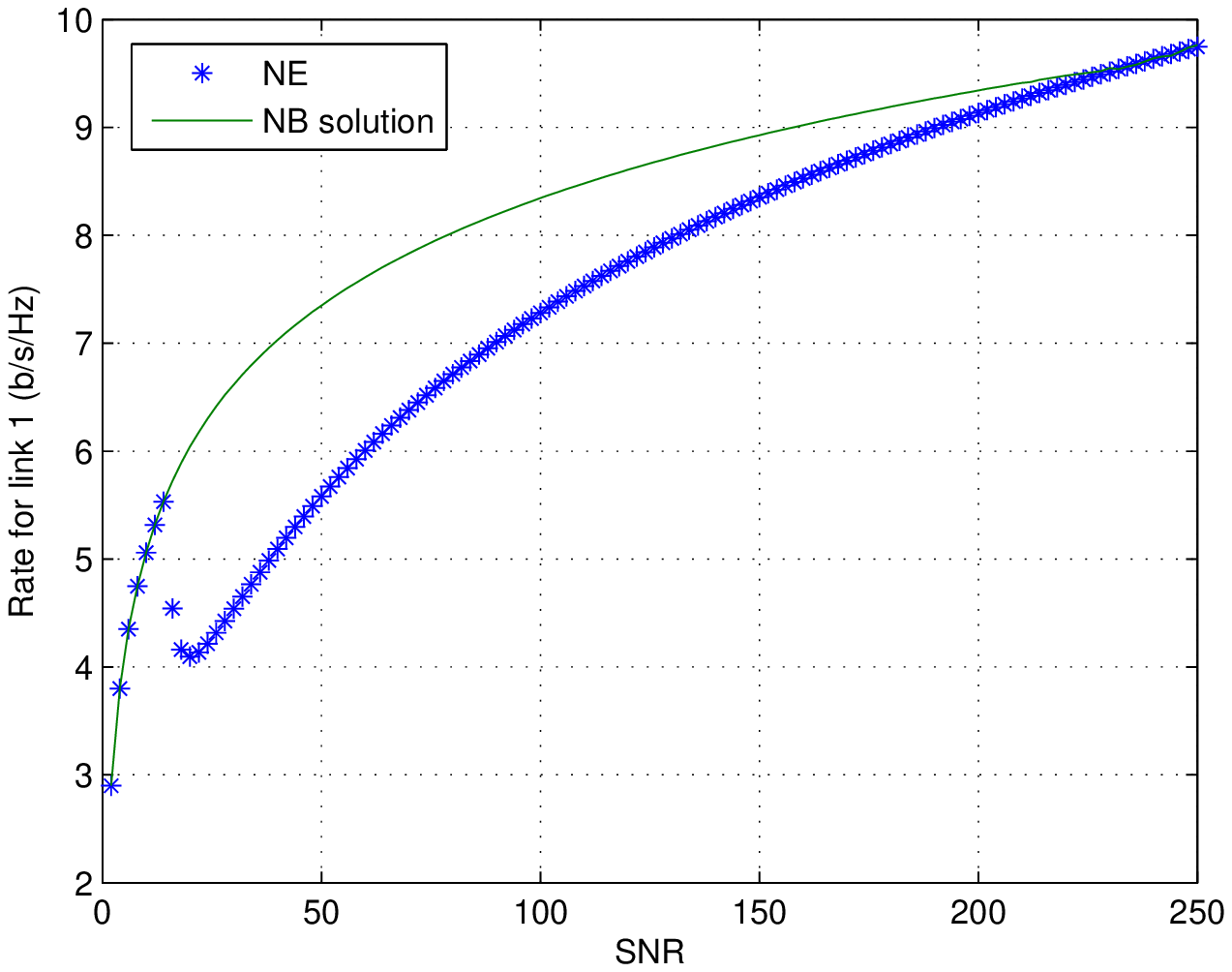} \\
\end{tabular}\\
\vspace*{-0.25cm}
\begin{center}
\small Fig.~6. Impact of the SNR on the NE and NB solution (INR=14).
\end{center}
\end{center}

\begin{center}
\begin{tabular}{c}
\hskip-0.6cm\epsfxsize=14cm\epsffile{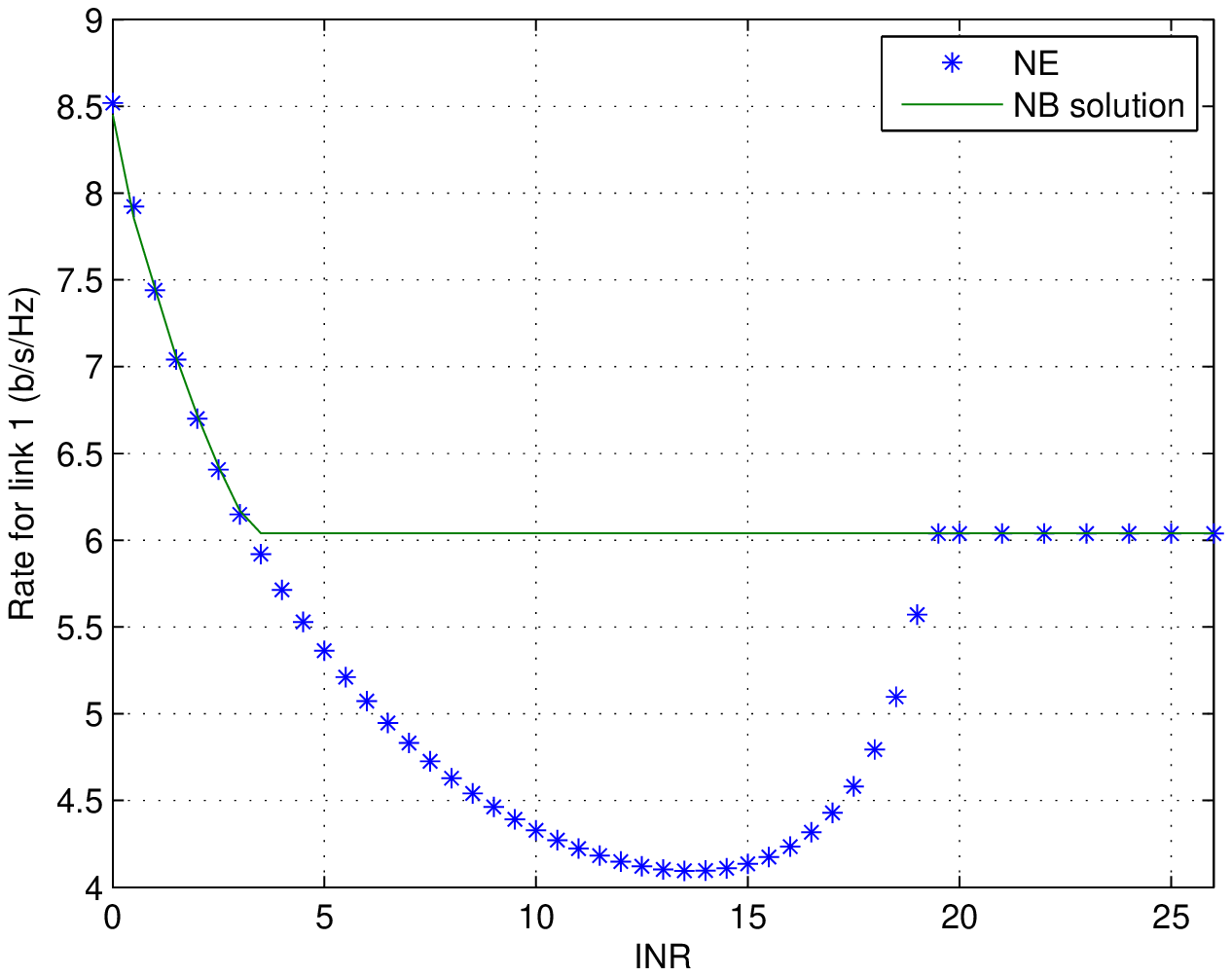} \\
\end{tabular}\\
\vspace*{-0.25cm}
\begin{center}
\small Fig.~7. Impact of the INR on the NE and NB solution (SNR=20).
\end{center}
\end{center}

\begin{center}
\begin{tabular}{c}
\hskip-0.6cm\epsfxsize=14cm\epsffile{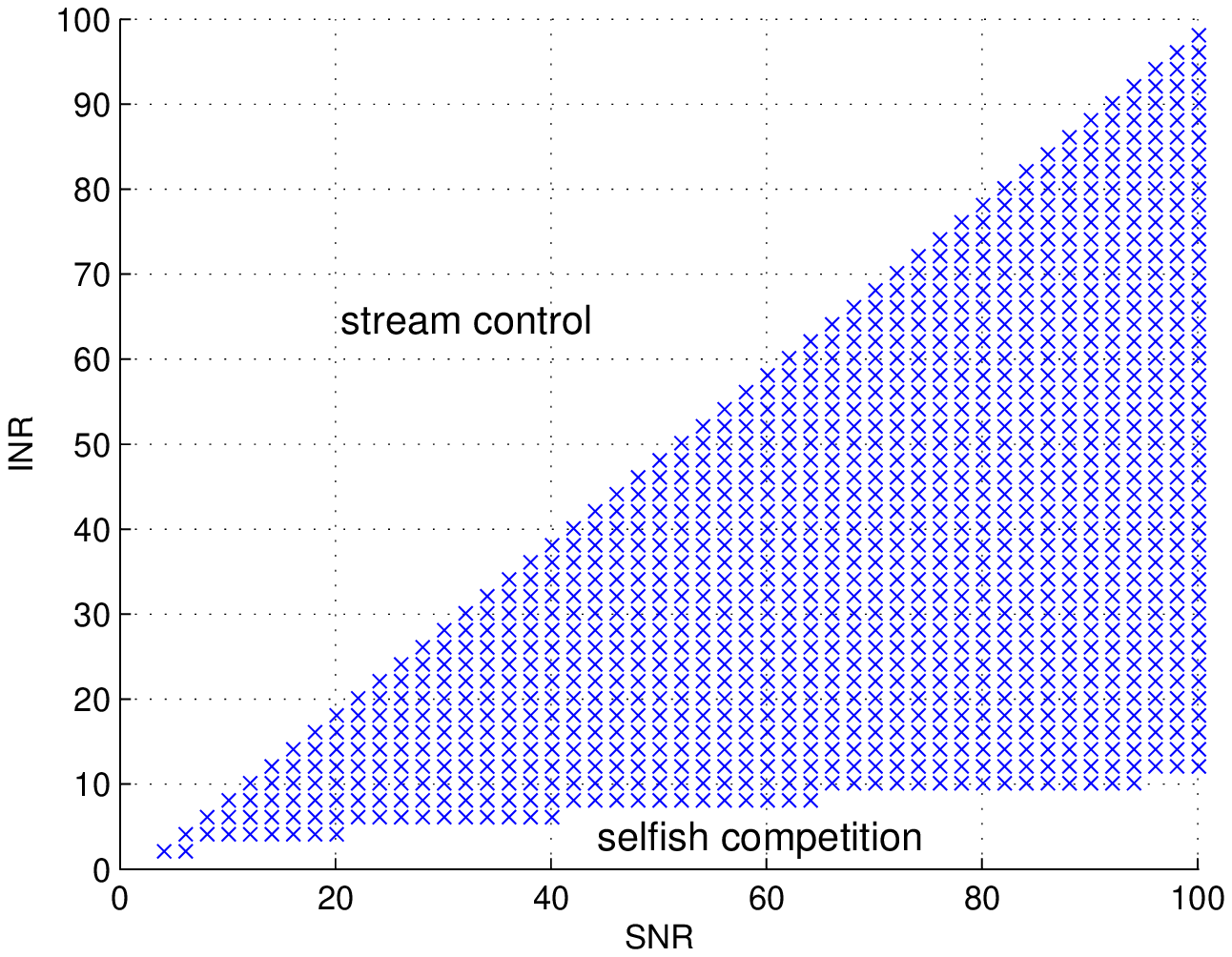} \\
\end{tabular}\\
\vspace*{-0.25cm}
\begin{center}
\small Fig.~8. Existence of the NB solution at different SNRs and INRs.
\end{center}
\end{center}

\begin{center}
\begin{tabular}{c}
\hskip-0.6cm\epsfxsize=14cm\epsffile{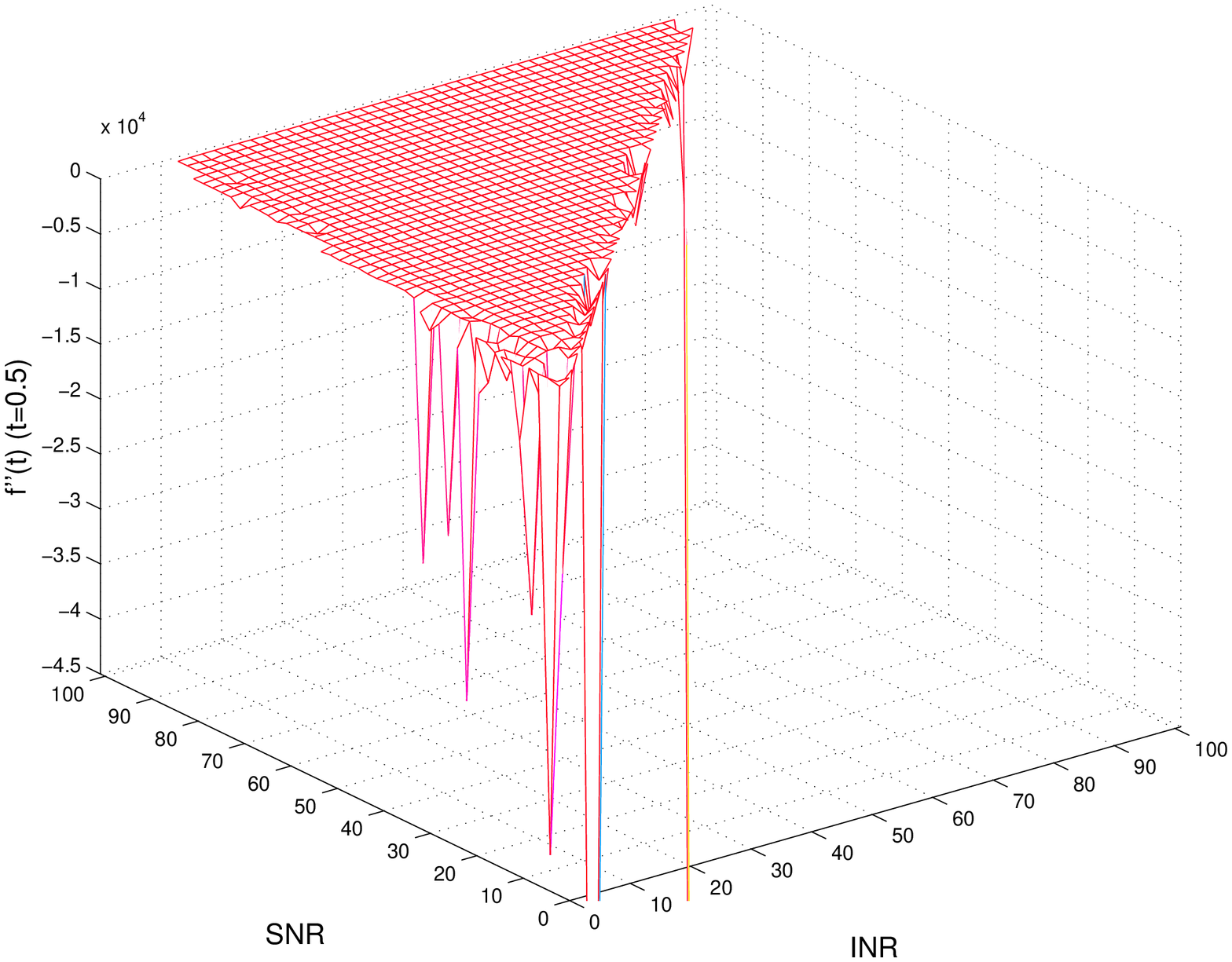} \\
\end{tabular}\\
\vspace*{-0.25cm}
\begin{center}
\small Fig.~9. One instance of $f^{''}(t)$ exemplifying the uniqueness of the NB solution.
\end{center}
\end{center}

\end{document}